\documentclass[reprint, amsmath,amssymb,onecolumn,aps]{revtex4-2}

\usepackage{graphicx}
\usepackage{dcolumn}
\usepackage{float,soul,caption,mathtools}
\usepackage{bm}

\newcommand{\be}{\begin{eqnarray}}
\def \bes {\begin{subequations} }
\def \ees {\end{subequations}}
\newcommand{\ee}{\end{eqnarray}}
 
\def \({\left(}
\def \){\right)}

\def\meref{Pradeep:2022eil}
\def\meappref{Karthein:2025hvl}

\newcommand\hT{{h_T}}
\newcommand\rT{{r_T}}
\newcommand\hmu{{h_\mu}}
\newcommand\rmu{{r_\mu}}
\newcommand\e{{\epsilon}}

\newcommand\bd{{\beta\delta}}

\newcommand\del{\partial} 

\usepackage[normalem]{ulem}
\usepackage{amsmath,caption,graphicx,subfigure,comment,multirow,xcolor,cancel,float,mathrsfs}

\usepackage{hyperref}

\begin{document}

\graphicspath{{{Plots/Alongfreeze-outCurve/},{Plots/ContourPlots/}}}

\preprint{APS/123-QED}

\title{Equation of state and cumulants of proton multiplicity in equilibrium near critical point from Pad\'e estimates}
\author{G\"ok\c ce Ba\c sar}
 \affiliation{Department of Physics, University of North Carolina at Chapel Hill, NC, 27599, USA}

\author{Maneesha Pradeep}
\affiliation{
 Centre for High Energy Physics, Indian Institute of Science, Bangalore, 560012, India
}%

\author{Mikhail Stephanov}
\affiliation{
 Department of Physics, University of Illinois at Chicago
}

\date{\today}

\begin{abstract} 
The fluctuations of proton multiplicity in heavy-ion collisions are the key observables in the search for the QCD critical point. In this work we present an approach to constraining the cumulants of proton number based on the analytical properties of the QCD equation of state in the vicinity of the  critical point. 
We show that, under the assumption of local equilibrium, the features of the collision energy dependence, such as the peaks and the dips of the cumulants, are significantly constrained by the data on the Lee-Yang singularity structure available via Pad\'e resummation of the lattice QCD data.
Furthermore, we identify four topologically distinct scenarios, all within the uncertainty range of the Pad\'e estimates for the non-universal mapping parameters, classified based on the location of the critical point and the slope of the chiral crossover curve with respect to the freeze-out curve. These different scenarios result in \textit{qualitatively} different critical signatures, especially for the third factorial cumulant and thus could be potentially discriminated using the experimental data. 
\end{abstract}

\maketitle

\section{Introduction}

Even after half a century since Quantum Chromodynamics (QCD) has been established as the microscopic theory of strong interactions, its phase diagram remains largely undetermined. This is due to its strongly interacting nature of the theory which renders perturbation theory inapplicable in the region where the temperature and baryon chemical potential are not asymptotically large. Numerical Monte Carlo calculations in lattice QCD are also severely limited at finite baryon density due to the sign problem \cite{deForcrand:2009zkb, Aarts:2023vsf}. An outstanding question regarding the phase diagram of QCD is whether there is a critical point at finite baryon chemical potential and temperature in the transition region  between the quark gluon plasma and hadron gas phases. 

The search for the critical point has been the driving motivation for the Beam Energy Scan program at the Relativistic Heavy Ion Collider as well as future experiments such as the Compressed Baryonic Matter at FAIR \cite{Stephanov:2004wx,CBM:2016kpk,Luo:2017faz,Busza:2018rrf,Bzdak:2019pkr,Luo:2022mtp}. The key signatures of the critical point that these experiments look for are associated with fluctuations. The magnitude of thermal fluctuations increase  near a critical point and as the fireball created in the collisions expand and cool down, the fluctuations freeze out and leave imprints on the spectrum of final state hadrons which the experiments measure \cite{Stephanov:1998dy,Stephanov:1999zu,Stephanov:2008qz,Stephanov:2011pb,An:2021wof,Du:2024wjm}. On the theory side, if the critical point exists it is located in the strongly interacting region inaccessible to analytical calculations. At the same time, based on symmetry considerations, if it exists, the critical point must be in the same universality class as the 3d-Ising model.
As a result the critical contribution to the QCD equation of state can be mapped to the Ising equation of state, whose relevant parameters are the reduced temperature, $r$, and the magnetic field $h$\cite{Nonaka:2004pg,Parotto:2018pwx,Pradeep:2019ccv,Kahangirwe:2024cny}.
The universality also determines the critical exponents. On the other hand, neither how the $r$ and $h$ parameters are mapped to the QCD parameters, $T$ and $\mu$, nor the location of the critical point can be determined via universality. They have to be extracted from QCD. Furthermore, {\em dynamics} of fluctuations also have to be taken into account in a quantitative description of the critical fluctuations that occur in the expanding fireball created in heavy ion collisions. A comprehensive theoretical framework that takes into account all these different factors is currently being developed.    

In this paper, we focus on the critical contribution to fluctuations in equilibrium. Our main goal is to relate the critical part of the equation of state to the experimental observables, which are the factorial cumulants of the proton multiplicity.
We use a lattice guided resummation method based on Pad\'e approximants to estimate the location of the critical point as well as the non-universal mapping parameters. The method we follow is to extract the location of complex Lee-Yang singularities of the QCD equation of state from the lattice data for different values of temperature\cite{Basar:2021hdf,Basar:2023nkp}. From this trajectory, it is possible to estimate the location of the critical point and constrain the non-universal mapping parameters. By using these estimates, we then compute the factorial cumulants of proton number multiplicities via the maximum entropy freeze-out method \cite{Pradeep:2022eil}. 

Our results show how the shape of the critical contribution to the cumulants as a function of collision energy depends on the non-universal mapping parameters and how the location of the complex singularities directly constrains this shape.  We shall also discuss implications of the interplay between the critical part of the equation of state and the hadronic freeze-out curve for the cumulants, both qualitatively and quantitatively. 

This paper is organized as follows. In Section \ref{Sec:generalities} we collect some observations regarding the QCD phase diagram which are relevant for this work. Most of the discussion in this section is not new. In Section \ref{Sec:Pade} we summarize the conformal Pad\'e approach to Lee-Yang singularities and list the parameter range for the critical point and mapping parameters that we use as input for our computation of the proton number cumulants. The next section, Sec. \ref{Sec:MEAndEoSDetails} is dedicated to the Maximum Entropy freeze-out method that relates the critical fluctuations in QGP to Hadron Gas. The results for the proton number cumulants are presented in Section \ref{Sec:Results}.

\section{General Observations on the Crossover/Phase Transition}
\label{Sec:generalities}

In this section we recapitulate certain properties of the phase diagram of QCD. This section is not meant to be a review of the QCD phase diagram; it is rather a collection of observations relevant for putting our results and their discussion into a context. Most observations are not knew, but their synthesis and some conclusions specific to the questions we address in this paper have not been presented, to our knowledge.

The two key non-perturbative phenomena that are relevant for the QCD phase diagram are chiral symmetry breaking and confinement. Chiral symmetry is probed by the chiral condensate $\sigma=\langle \bar q q\rangle$ which is nonzero in the broken phase at low temperature and zero in the symmetric phase at high temperature. Confinement, on the other hand, cannot be defined or quantified as straightforwardly \cite{Greensite:2003bk}. Setting aside the challenges, a commonly used ``order parameter" for confinement is the Polyakov loop, $L$, whose expectation value is proportional to the exponent of the free energy $F$ of a single static quark: $\langle L\rangle\sim\exp(- F/T)$. 
The associated symmetry is the center symmetry of the gauge group. The confined phase occurs at low temperatures where the center symmetry is restored, $\langle L\rangle=0$, and the static quark free energy is infinite; whereas in the deconfined phase, occurring at high temperature, the symmetry is broken, $\langle L\rangle\neq0$, and the free energy is finite. This interpretation warrants some caution. Polyakov loop is not really an order parameter in the conventional sense because it is non-local and is only defined in Euclidean space. Furthermore, there are non-Abelian gauge theories without any center that exhibit phenomena akin to confinement. Nevertheless, keeping these caveats in mind, Polyakov loop can still be interpreted as an indicator of confinement. At physical values for the quark masses, neither chiral nor center symmetry are exact symmetries of QCD and consequently there are no order parameters that one can use to sharply define or distinguish different phases.

\begin{figure}
\center
\includegraphics[scale=0.5]{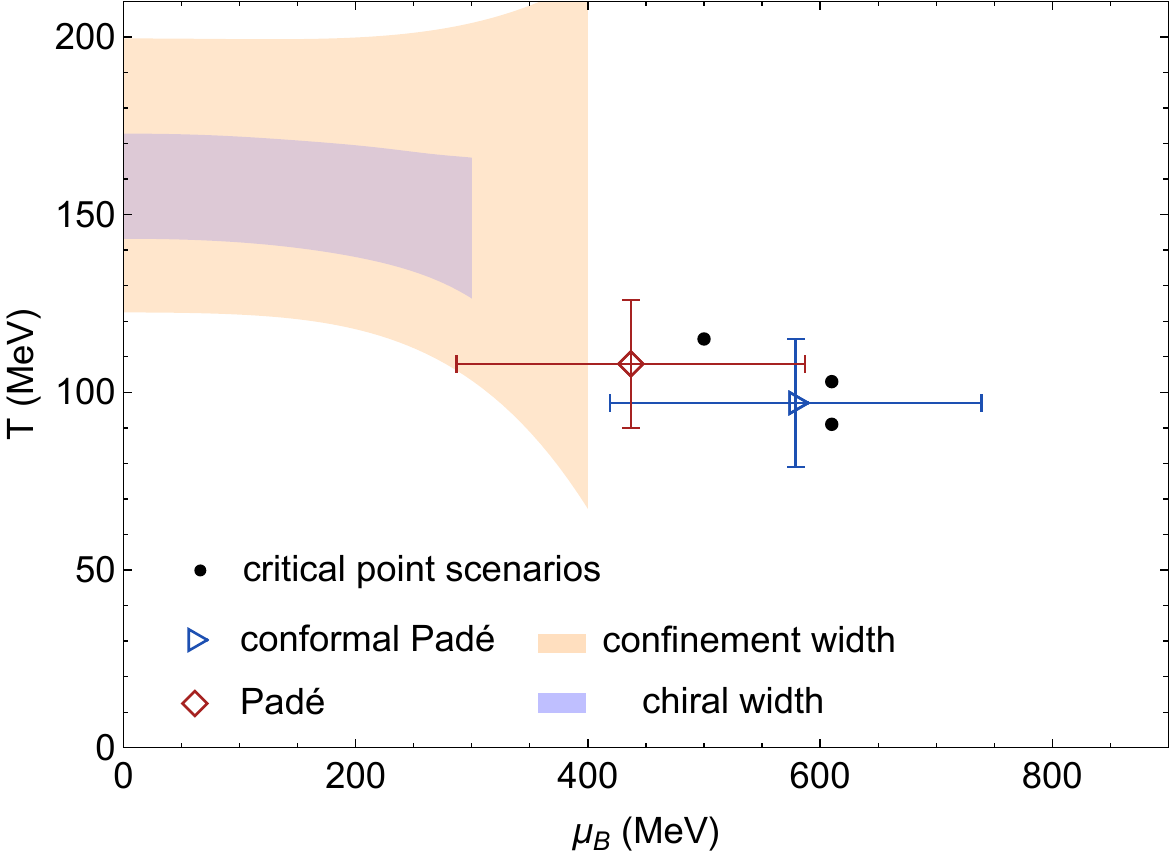}
 \captionof{figure}{Estimated widths of confinement and chiral transitions from the lattice \cite{borsanyi2024qcddeconfinementtransitionline}.  The two estimates based on Pad\'e resummation for the critical point are shown in red and blue. The black dots indicate the three exemplary locations for the critical point we choose (see Table \ref{table:parameters-Pade}).}
\label{Fig:pd}
\end{figure}

At zero baryon density, it is now well established from the lattice studies that the thermodynamic properties vary smoothly with temperature indicating a crossover rather than a sharp phase transition. Likewise, the chiral condensate decreases smoothly with increasing temperature, and the chiral susceptibility, $\chi=\partial\sigma/\partial m_q$, peaks in the $T=155-160$ MeV range\cite{Bazavov_2019,Borsanyi_2020}. The typical width of this chiral crossover is approximately $30$ MeV. The deconfinement transition is even harder to pin down. The Polyakov loop smoothly changes from smaller to larger values with increasing temperature and shows signs of saturation at temperatures close to 1 GeV. However, the change occurs in a much wider range than the chiral transition  without any indication of an inflection point in the chiral crossover regime \cite{Bazavov_2013}. In practice, the Polyakov loop is plagued by UV effects and as a result, other ``healthier" observables derived from the Polyakov loop, such as the static quark free energy \cite{Bazavov_2013}, static quark entropy \cite{borsanyi2024qcddeconfinementtransitionline}, or 't Hooft loop \cite{Dumitru_2011} have been used to probe the confinement transition instead. They do indicate signs of a confinement crossover in the chiral crossover regime, albeit wider. For example, a state-of-the art computation of the static quark entropy estimates the width to be approximately $80$ MeV -- much larger than the chiral crossover width of $30$ MeV (see Fig \ref{Fig:pd}) \cite{borsanyi2024qcddeconfinementtransitionline}. The significant difference between the chiral and deconfinement widths is not surprising. Chiral and center symmetries become exact in opposite limits of QCD with respect to quark mass.  The chiral condensate is a good order parameter when the quark masses go to zero whereas the Polyakov loop is a good ``order parameter" (with the caveats mentioned above) when the quark masses go to infinity. For physical quark masses, the chiral symmetry is closer to being exact than the center symmetry leading to a smaller width in $\chi$ . Not surprisingly, the difference becomes more dramatic for smaller quark masses. In fact, lattice computations at smaller-than-physical quark masses show no inflection in the Polyakov loop whatsoever in the chiral crossover regime \cite{Clarke:2019tzf}. 
In Fig. \ref{Fig:pd} we show a recent set of estimations for the chiral and confinement widths from \cite{borsanyi2024qcddeconfinementtransitionline}.

For nonzero baryon density, it is widely believed that the smooth crossover turns into a first-order transition for some value of baryon chemical potential with a second-order critical point marking the beginning (or the end, depending on the point of view) of the first-order curve. At the critical point, both the chiral and Polyakov loop susceptibilities are expected to diverge. The state-of-the art lattice calculations indicate the absence of a critical point in the range for $\mu_B\lesssim 450$ MeV \cite{borsanyi2025latticeqcdconstraintscritical}. Due to the sign problem, it is extremely challenging to push this bound further.

Despite the sign problem, it is possible to estimate the location of the critical point with different methods such as Pad\'e type resummations of the lattice results \cite{Bollweg_2022,Basar:2023nkp,Clarke_2024}, Dyson-Schwinger equations \cite{Gao:2020fbl,Gunkel:2021oya}, functional renormalization group \cite{Fu_2020}, constant entropy contours \cite{shah2024locatingqcdcriticalpoint}, and holography \cite{Hippert_2024}. Remarkably, all of these different approaches hint towards the existence of a critical point within the region $\mu_B\in (350,700)$ MeV, $T_c \in (80,120)$ MeV. Notably, most of these results curiously accumulate around a central value of $\mu_c\sim 600$ MeV.  In this paper, we shall explore the implications of  the Pad\'e based estimations for the critical contribution to the QCD equation of state. Results from two different type of Pad\'e estimates along with three different locations for the critical point in the Pad\'e uncertainty region we consider in this paper are shown in the left panel of Fig. \ref{Fig:pd}. 

Considering the possibility of a critical point around $(\mu_c,T_c)=(600,100)$ MeV raises several interesting questions. First of all, we expect the chiral crossover curve, defined via the peak of chiral susceptibility, to go smoothly from the so-called pseudo-critical point, $(\mu_B,T)\approx(0,158)$ MeV, to the critical point where it turns into a first-order curve. This first order curve most  likely terminates at zero temperature for some value of the baryon chemical potential, $\mu_0>\mu_{LG}=922$ MeV\footnote{This value corresponds to the end point of the nuclear liquid-gas transition $\mu_{LG}=m_p+\Delta E_{bind}\approx922$MeV at zero temperature. The chiral transition supposedly occurs at a higher density} with its slope approaching negative infinity. 
The vertical approach to $T=0$ point is a combination of Nernst theorem which indicates that the entropy vanishes at $T=0$, and Clausius-Clyperon relation which implies the slope is inversely proportional to the discontinuity in entropy which goes to zero \cite{Halasz:1998qr}. At the same time, from the lattice computations, we know that the crossover curve around $\mu=0$ is relatively flat with the curvature being small and the coefficient of the quartic term being consistent with zero \cite{Bazavov_2019,Borsanyi_2020}. As a result, the existence of a critical point at a relatively large chemical potential such as $\sim600$ MeV puts a constraint on the shape of the crossover/first-order curve. To satisfy the asymptotic behaviors at $\mu_B=0$ and $T=0$, and pass through the critical point at the same time, the transition curve has to bend towards the chirally symmetric phase, possibly with an inflection point at large $\mu$. A possible, perhaps natural, way how this could happen is that the curvature of the chiral transition curve develops a discontinuity at the critical point. The slope is continuous at the critical point and is equal to the slope of the $h=0$ line on the phase diagram. However, the curvature can be discontinuous; the crossover and first-order sides could have different curvatures. In Appendix \ref{Sec:appendix-GL}, we show that this is in fact what happens in the mean field theory, which is applicable near the tricritical point at $m_q=0$. Furthermore, the discontinuity is such that the curve bends towards the symmetric phase in the first-order side, consistent with our expectation. To our knowledge, the existence of such a curvature discontinuity away from the chiral limit and its implications have not been discussed in the literature.

Secondly, the location of the critical point and the slope of the first order curve with regards to the experimental freeze-out curve for the heavy ion collisions leads to topologically distinct scenarios with qualitatively different features that can be experimentally distinguished. We discuss these scenarios in detail in Sec. \ref{SubSec:freeze-outScenarios}.

\section{Results from the lattice and Pad\'e resummations}
\label{Sec:Pade}

Critical point is a singularity of the equation of state. This means that Taylor expansion breaks down at this point. The singularity in pressure, $P^{\rm singular}$, appears in such a way that the derivatives beyond the first order do not exist at the critical point, i.e.,
\begin{equation}
    P(\mu,T)-P(\mu_c,T_c)-n_c(\mu-\mu_c)-s_s(T-T_c) = P^{\rm singular}(\mu,T) + \mathcal O(\mu-\mu_c,T-T_c)^2\,,
\end{equation}
where $n_c$ and $s_c$ are, respectively, the baryon and the entropy densities, $n=\partial P/\partial\mu$, $s=\partial P/\partial T$, evaluated at the critical point. In other words, the susceptibilities (second order derivatives of pressure) are divergent at the critical point.

The singularity is universal. This means that at the QCD critical point, which is in the universality class of the Ising model, the {\rm leading} singular contribution can be expressed in terms of the leading singular contribution to the Ising model equation of state via the following mapping:
\be
\label{Eq:Psing}
 P^{\text{singular}}(\mu,T)=-T^{4}_c\, G_{\rm{Ising}}(r(\mu,T)\, h(\mu,T))
\ee
where $G_{\rm{Ising}}$ is the leading singular contribution to the Ising model equation of state in terms of the reduced temperature $r$ and ordering field $h$ with the critical point located at $r=h=0$.
The equation of state of the Ising model, $G_{\rm Ising}(r,h)$ is not known analytically. Instead we use the standard parametric representation introduced in Ref. \cite{Guida:1996ep}. The same parameterization has been used in Refs. \cite{Parotto:2018pwx,Kahangirwe:2024cny}. 

Near the critical point, the Ising variables, $r$ and $h$, can be mapped to the temperature and baryon chemical potential of QCD via a linear map \cite{Parotto:2018pwx,Pradeep:2019ccv}:
\begin{eqnarray}
\label{Eq:map-linear}
\begin{pmatrix}
r\\ h
\end{pmatrix}=
\begin{pmatrix}
r_T &&r_\mu 
\\
h_T   && h_\mu 
\end{pmatrix}
\begin{pmatrix}
T-T_c \\ \mu-\mu_c
\end{pmatrix}
=\frac{1}{  T_c w\rho \sin\alpha_{12}}\begin{pmatrix}
\cos\alpha_2 && \sin \alpha_2
\\
-\rho \cos\alpha_1   && -\rho \sin\alpha_1
\end{pmatrix}
\begin{pmatrix}
T-T_c \\ \mu-\mu_c
\end{pmatrix}
\quad
\end{eqnarray}
 The two angles in this map, $\alpha_1$ and $\alpha_2$, respectively determine direction of the $h=0$ and $r=0$ axes in the $T,\mu$ plane. We define 
 \begin{equation}
    \alpha_{12}\equiv \alpha_1-\alpha_2\,.
    \label{Eq:alpha12}
\end{equation}
 The slope of the crossover/first-order transition line at the critical point is $\tan\alpha_1$. The parameters $w$ and $\rho$ parameterize the size and the shape of the critical region. In particular, larger values of $\rho$ or smaller values of $w$ make the critical region larger. We stress that none of these parameters that quantify the mapping are universal unlike the critical exponents. They have to be computed directly from QCD just like the location of the critical point, $T_c$ and $\mu_c$. It is possible to express these parameters in terms of combinations of various derivatives of pressure with respect to $T$ and $\mu$ \cite{Pradeep:2019ccv}. Of course, directly computing them on the lattice faces the sign problem. 

It is possible, however, to extrapolate the lattice data for small $\mu\equiv\mu_B$\footnote{In this paper we will use $\mu$ and $\mu_B$ interchangeably to denote the baryon chemical potential.} to higher values of $\mu_B/T$ using various resummation techniques. In this paper, we shall focus on an extrapolation scheme based on Pad\'e approximants. The starting point in this framework is that the QCD equation of state, in general, has singularities in the complex $\mu_B$ plane \cite{Stephanov:2006dn}. The two complex conjugate singularities which at $T=T_c$ meet at the critical point on the real $\mu_B$ axis are 
known as the Lee-Yang (LY) edge singularities. At fixed $T>T_c$ and in the vicinity of each of the edge singularities, $\mu_{\rm LY}$, the equation of state has a singular contribution :
 \begin{equation}
    P(\mu)-P(\mu_{\rm LY})-n(\mu_{\rm LY})(\mu-\mu_{\rm LY})\sim (\mu-\mu_{\rm LY})^{1+\sigma}
 +\mathcal O(\mu-\mu_{\rm LY})^2\quad\mbox{as $\mu\to\mu_{\rm LY}$}\,.
 \end{equation}
A recent numerical calculation gives $\sigma=0.0742(56)$ at d=3 \cite{An:2016lni}. The location of the edge singularities depend on the temperature, generating what we call the ``Lee-Yang trajectory", denoted by $\mu_{\rm LY}(T)$. Since the equation of state is real for real values of $\mu$, the singularities come in complex conjugate pairs. 

At the critical point the complex conjugate LY singularities meet on the real axis \cite{Yang:1952be,Lee:1952ig}:
\begin{equation}
    \mu_{\rm LY}(T_c)=\mu_c
\end{equation}
where $T_c$ and $\mu_c$ are the critical temperature and critical baryon chemical potential as usual. In the vicinity of the critical point, the behavior of the LY trajectory can similarly be mapped to the Ising model where they take the form 
\begin{equation}
h r^{-\bd}=\pm ix_{\rm LY}\,.
\label{Eq:LY-ising}
\end{equation}
Here $x_{\rm LY}$ is a numerical constant whose value has been computed via various methods \cite{Rennecke:2022ohx,Connelly:2020gwa,Johnson:2022cqv,Karsch:2023rfb}. We will use the value $x_{\rm LY}\approx 0.246$. Finally, expressing Eq. \eqref{Eq:LY-ising} in terms of the QCD variables via the linear map given in Eq. \ref{Eq:map-linear}, we arrive at the following trajectory of singularities as a function of temperature in the critical region \cite{Stephanov:2006dn}
\begin{eqnarray}\label{Eq:ly-traj}
\mu_{\rm LY}(T)\approx\mu_c -c_1(T-T_c) \pm i x_{\rm LY} c_2 (T-T_c)^\bd, 
\\
\text{where }c_1={\frac\hT\hmu}=\tan\alpha_1 \quad c_2=\frac{\rmu^\bd}{\hmu} \left(\frac\rT\rmu-\frac\hT\hmu \right)^\bd\,.
\end{eqnarray}

The form of this trajectory is fixed by the universality. However the coefficients $c_1$ and $c_2$ are non-universal.  In Refs. \cite{Bollweg_2022,Basar:2023nkp,Clarke_2024} the value of $\mu_{\rm LY}$ for different temperatures has been extrapolated from the lattice data by using Pad\'e resummation as we shall detail further shortly. From these estimates, it is possible to calculate $T_c$, $\mu_c$ $c_1$ and $c_2$ which are then used as inputs for calculating the proton number cumulants.  The critical temperature is determined by ${\rm Im} \mu_{\rm LY}(T_c)=0$. In its vicinity, to leading order, the real part of the trajectory depends linearly on $T-T_c$ where  the slope and the intercept are equal to the critical chemical potential, $\mu_c$, and the crossover slope, $-\tan\alpha_1$, respectively. The imaginary part of the trajectory, likewise, contains information about the mapping parameters.  By combining equations \eqref{Eq:map-linear} and \eqref{Eq:ly-traj} we can write
\begin{equation}
    c_2=\frac{1}{T_c^{\bd-1} w^{\bd-1} \rho^\bd }\frac{|\sin\alpha_{12}|}{|\sin\alpha_1|^{\bd+1}}\,.
\end{equation}
As a result, by calculating $c_2$ via Pad\'e or some other way, we can use this equation to constrain the mapping parameters. For instance, we can express $\rho$ as a function of the remaining parameters as
\begin{eqnarray}
\label{Eq:rho}
\rho(w,\alpha_1,\alpha_2)=\left(\frac{1}{c_2 T_c^{\bd-1} w^{\bd-1}}\frac{|\sin\alpha_{12}|}{|\sin\alpha_1|^{\bd+1}} \right)^{1/\bd}\,.
\end{eqnarray}
Furthermore, instead of $\rho$, it is actually better to work with the rescaled parameter
\begin{equation}
    \bar\rho\equiv \rho w^{1-1/\bd}\,.
    \label{Eq:rho-bar}
\end{equation}
This combination is rather special. We show in Sec \ref{Sec:EoS} that in the critical region, the shape of the proton number factorial cumulants depend on $\bar\rho$,  but not $w$. The parameter $w$ instead changes the overall scale of the factorial cumulants preserving the shape. By using  Eq. \eqref{Eq:rho} we can express $\bar\rho$ as
\begin{eqnarray}
\label{Eq:rho-bar-Pade}
\bar\rho=\left(\frac{1}{c_2 T_c^{\bd-1} }\frac{|\sin\alpha_{12}|}{|\sin\alpha_1|^{\bd+1}} \right)^{1/\bd}
\end{eqnarray}
Therefore we arrive at a remarkable result; up to the orientation of the $r=0$ axis controlled by $\alpha_2$, the imaginary part of the Lee-Yang trajectory  fixes the shape of the factorial cumulants of proton multiplicity in the critical region. 

We now turn to the Pad\'e resummation scheme. The basic strategy is to first estimate the location of the nearest singularity to origin in the complex $\mu$ plane. The truncated Taylor series for the equation of state computed on the lattice,
\begin{equation}
    \frac{p(T,\mu_B)-p(T,0)}{T^4}\approx \sum_{n=1}^N \frac{\chi_{2n}(T)}{(2n)!} \left( \frac{\mu_B}{T} \right)^{2n}, 
    \label{Eq:p-Taylor}
\end{equation}
does not have any singularities, as it is a polynomial. A standard way of extracting singularities from such a form is to express it as rational function whose coefficients are chosen to match the original Taylor coefficients:
\begin{eqnarray}
    p(T\,\mu_B)\approx\frac{q_1(\mu_B^2)}{q_2(\mu_B^2)}\,.
    \label{Eq:p-Pade}
\end{eqnarray}
Here $q_1$ and $q_2$ are polynomials of order $N/2$ whose coefficients are fixed in terms of $\chi_{2n}(T)$s in Eq. \eqref{Eq:p-Pade} upon expansion. This form is known as the Pad\'e approximant. It is known that the zeroes and poles of the Pad\'e approximant accumulate around the original singularity of the underlying function, $p(T,\mu_B)$ in our case. The estimation of the singularity can be improved by performing the Pad\'e resummation after a conformal map, 
\begin{equation}
    \mu_B^2\equiv \phi(\zeta)
\end{equation}
that maps the complex $\mu_B^2$ plane into another domain, typically a compact one such as the unit disk, in a way that takes into account the fact that there are two branch points in the complex $\mu_B$ plane and then do the Pad\'e resummation in the $\zeta$ plane:
\begin{equation}
    p(T,\mu_B)\approx \frac{\tilde q_1(\zeta)}{\tilde q_2(\zeta)}\big|_{\zeta=\phi^{-1}(\mu_B^2)}\,.
    \label{Eq:conf-Pade}
\end{equation}
The types of conformal maps that are proven to approximate the original function more accurately than ordinary Pad\'e approximants are known \cite{Costin:2020pcj}.  These conformal maps also push the regime of applicability of the resummation beyond the radius of convergence, even beyond the first Riemann sheet \cite{Basar:2021gyi,Basar:2021hdf,Basar:2023nkp}. In this work we consider the results from two different conformal maps. The image of the singularities in the $\zeta$ plane appear as poles and zeroes of the polynomials $\tilde q_{1,2}(\zeta)$ in Eq.\eqref{Eq:conf-Pade} which can be mapped back to the $\mu_B^2$ plane via $\phi$. 

The next step is to use this method to estimate the singularities for different values of $T$ and construct the LY trajectory. From the trajectory one can determine the location of the critical point, from the relation Re $\mu_{\rm LY}(T=T_c)=\mu_c$. Finally, using the scaling form given in Eq. \eqref{Eq:ly-traj} as a fit form, one can also read off $c_1$ and $c_2$ from the real and imaginary parts of the Lee-Yang trajectory  The first coefficient $c_1=\tan\alpha_1$ fixes the slope of the transition curve at $T_c$. Along with the value of $c_2$ also fixed from Pad\'e estimate, Eq. \eqref{Eq:rho-bar-Pade} defines a one-parameter family of $\bar\rho$s, parameterized by $\alpha_2$. In this work we treat $\alpha_2$ as a free parameter and plot the proton number cumulants for a select exemplary set of $\alpha_2$s. It is worth noting that the value of $\bar\rho$ is independent of $\mu_c$.  

The results of Pad\'e resummation based on Ref. \cite{Basar:2023nkp} is given in Table \ref{table:Pade}. Here we quote results from three different resummation methods; ordinary Pad\'e and Pad\'e paired with two different conformal maps \cite{Basar:2021hdf,Basar:2021gyi}. Ordinary Pad\'e refers to usual Pad\'e resummation: we first express the truncated Taylor series approximation of the equation of state, calculated on the lattice. For the Taylor coefficients we refer to the recent results by the HotQCD collaboration, where the equation of state is calculated up to order $\mu_B^8$ \cite{Bollweg_2022}. Given that the equation of state is a function of $\mu_B^2$, there are 4 terms in the Taylor series, and a $[2,2]$ Pad\'e approximant was used.  From the results in Table \ref{table:Pade}, one can see that despite their different functional forms, the two different conformal maps lead to similar results, which slightly differ from ordinary Pad\'e. In Fig. \ref{Fig:pd} we show the location of critical point obtained from ordinary Pad\'e and only one conformal Pad\'e (that uses the so-called uniformizing map), since the other map leads to a similar result and we do not want to make the figure overly crowded. We note that the errors quoted in Table \ref{table:Pade} and in Fig. \ref{Fig:pd} are statistical errors inherited from the lattice. They include neither the lattice systematics nor the  Pad\'e errors due to the small number of Taylor coefficients. 

\begin{table}[h]
\renewcommand{\arraystretch}{1.7}{\begin{tabular}{  | c  | c | c | c |  c | } 
 \hline
 resum. method (conformal map) \, & $T_c$ (MeV) & $\mu_c $ (MeV) & crossover slope ($\alpha_1$) & $c_2 $ (MeV$^{1-\beta\delta}$)
 \\
\hline
uniformizing map \,&\,  $97^{+18}_{-18}$  \,&\,  $ 579^{+172}_{-160}$  \,&\,    $9.40^\circ\,^{+3.89}_{-3.81}$ \,&\,  $2.22^{+0.52}_{-0.86}$ \,
\\  
 two-cut map\,&\, $ 100^{+18}_{-18}$   \,&\,  $557^{+175}_{-150}$  \,&\,   $8.69^\circ\,^{+3.91}_{-3.83}$ \,&\, $2.56^{+0.58}_{-1.21}$ \,
\\  
ordinary Pad\'e (no conf. map) \,&\,  $108^{+21}_{-21}$  \,&\,  $ 437^{+114}_{-50}$  \,&\,    $4.55^\circ\,^{+3.41}_{-3.37}$ \,&\,  $3.35^{+0.82}_{-1.37}$ \,
\\
 \hline
 \end{tabular}
 }
 \caption{The Pad\'e estimates of the critical point and the mapping parameters from Eq. \eqref{Eq:ly-traj} with $1\sigma$ uncertainty.}
\label{table:Pade}
\end{table}

\begin{figure}
\center
\includegraphics[scale=0.5]{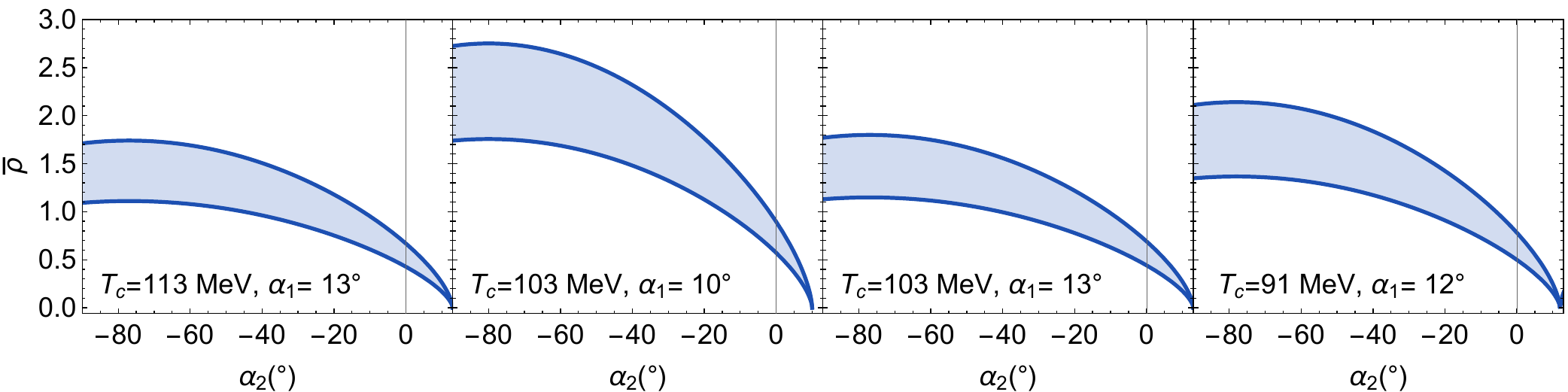}
 \captionof{figure}{The constraint on $\bar\rho(\alpha_2)$ for the temperatures we use in the critical point scenarios. The band stems from the uncertainty in $c_2$ given in Table \ref{table:Pade}.}
\label{Fig:rhobar-vs-alpha2}
\end{figure}

Equipped with the estimates for the critical point as well as the non-universal mapping parameters listed in Table \ref{table:Pade}, we calculate the singular, critical contribution to the equation of state. From this computation we  then extract the proton number fluctuations using maximum entropy freezeout approach described in the next Section. In principle it is possible to take into account the uncertainties in these values and do a comprehensive Bayesian analysis to find the best description of the experimental data given these constraints.  Leaving such an extensive analysis to future, we follow a simpler strategy. We choose a set of exemplary parameters, constrained by the Pad\'e results and compute the proton number cumulants using this set. These values are chosen to illustrate that even the qualitative features of the cumulants depend sensitively on these non-universal parameters. The values we consider are summarized in Table \ref{table:parameters-Pade} and Fig. \ref{Fig:rhobar-vs-alpha2}

\begin{table}
\renewcommand{\arraystretch}{1.7}{\begin{tabular}{  | c  | c | c | c |  } 
 \hline
$\mu_c,T_c$ (in MeV)  &  536, 113  & 610, 91 & 610, 103
 \\
\hline
$\alpha_1$  \,&\, $13^\circ$  \,&\, $12^\circ$ \,&\,  $10^\circ,13^\circ$  \,
\\  
$\alpha_2$ \,&\, $0^\circ,-20^\circ,-89^\circ$  \,&\, $0^\circ,-6^\circ,-89^\circ$\,&\, $5^\circ,-6^\circ,-89^\circ$ \,
 \\
$w$ \,&\, 5  \,&\,  5  \,&\,  5  \,
 \\
 $\bar\rho$ \,&\, in Fig. \ref{Fig:HydroColdWithCrossing}  \,&\,  in Fig. \ref{Fig:HydroColdWithoutCrossing}   \,&\,  in Figs. \ref{Fig:HydroHotWithoutCrossing},\ref{Fig:HydroHotWithCrossing}  \,
\\
\hline
 \end{tabular}
 }
 \caption{The exemplary parameter choices for the critical point scenarios constrained by Pad\'e results.}
 \label{table:parameters-Pade}
\end{table}

\section{Maximum Entropy Approach to Multiplicity Fluctuations}
\label{Sec:MEAndEoSDetails}
\subsection{Maximum entropy freeze-out  for hydrodynamic fluctuations}
\label{Sec:ME}
In heavy-ion collision experiments, the primary observables are the multiplicities 
of final state hadrons. In order to relate the physics of hot QCD, such as the location and the properties of the critical point, to experimental measurements we need a theoretical framework that relates the hydrodynamic observables that describe the hot QCD plasma to the observables of the final state hadrons. Empirically, the mid-rapidity yields of various hadrons agree with a model that assumes that the entropy of the hadron-resonance gas ensemble is maximized at a particular instant of the hydrodynamic evolution, called chemical freeze-out, where the average particle multiplicities become frozen\cite{Andronic_2018}. Recently, this maximum-entropy freeze-out framework has been generalized to describe the event-by-event fluctuations of particle multiplicities \cite{\meref}, and implemented for strongly interacting matter in equilibrium traversing the vicinity of a 3D Ising-like critical point \cite{\meappref}. The idea in a nutshell is to find the most likely phase space distribution of the hadrons after freeze-out by maximizing the entropy under the set constraints that the average values of all conserved quantities as well as their correlators in the hadron gas match the values of the corresponding hydrodynamic correlators right before freeze-out. 

In this work, we employ the following strategy. We first constrain the location of the critical point, the crossover slope, as well as the scale factor $\bar{\rho}$ using improved Padé estimates. We then use the maximum entropy freeze-out framework to examine how these constraints affect the critical contribution to the factorial cumulants of proton multiplicities along the freeze-out curve, under the assumption of thermal equilibrium. We note that our framework is ``modular" in the sense that the maximum-entropy freeze-out step can be applied to other methods of extracting the critical equation of state. In other words, from the perspective of the maximum entropy freeze-out the location of the critical point and the mapping parameters are free inputs. We used the Pad\'e approach to constrain them in this work.   

We now briefly summarize the key ingredients of the maximum-entropy freeze-out framework for hydrodynamic fluctuations which we use to convert the hydrodynamic cumulants to factorial cumulants of proton multiplicity \cite{\meref}. We first set notation and conventions that we use in the remainder of the paper. We denote the distribution function of hadrons in phase space by $f_A$. Capital letter indices, such as $A$, label hadronic one-particle states:
\begin{equation}
    A\equiv\{\bm x,\bm p, m,q,\dots\}\,,
\end{equation}
referring to the position and momenta ${\bm x}$, ${\bm p}$ along with the energy (mass) $m$, (baryon) charge $q$, and any other quantum numbers this state might carry. The deviation from the average distribution is denoted as
\begin{equation}
    \delta f_A\equiv f_A- \langle f_A \rangle \,. 
\end{equation}
The central objects of interest are the connected correlators of hadrons,
\begin{equation}
    G_{A_1\dots A_k}\equiv \langle \delta f_{A_1}\dots\delta f_{A_k}\rangle_{\rm connected},
    \label{Eq:Gdef}
\end{equation}
whose integrals over the phase space are related to experimental observables such as multiplicity distributions and its cumulants. 

We denote the hydrodynamic densities of conserved quantities as $\psi_a$.
Here the lowercase subscripts/superscripts represent the type of  conserved quantity, such as energy density or baryon density, along with the spatial point it is evaluated at:
\begin{equation}
    a\equiv\{\e,\bm x\}, \{n,\bm x\},\dots\,.
\end{equation}
For example, in this notation the energy density at point ${\bm x}$ would be denoted as $\e({\bm x})=\psi_{\{\e,{\bm x}\}}$. The hydrodynamic densities exhibit thermal fluctuations as usual. We denote these deviations from the average as
\begin{equation}
    \delta \psi_a\equiv \psi_a- \langle \psi_a \rangle  \,.
\end{equation}
The connected hydrodynamic correlators of conserved densities are denoted by the letter $H$,
\begin{equation}
    H_{a_1\dots a_k}\equiv\langle\delta\psi_{a_1}\dots\delta\psi_{a_k}\rangle_{\rm connected}\,.
    \label{Eq:Hdef}
\end{equation}
The local conservation law implies that for every cell we have 
\be\label{Eq:cons}
	\psi_a= f_A P^{A}_{a}
	\ee	
where, for $A=\{\bm x,\bm p,m,q\}$ and $a=\{\epsilon,\bm y\}$ or $\{n,\bm y\}$,
\be \label{Eq:P}
P_{\{\e,\bm y\}}^{\{{\bm x},{\bm p},m,q\}}\equiv \sqrt{{\bm p}^2+m^2}\, \delta^3({\bm x}-{\bm y})\, , \qquad P_{\{n,\bm y\}}^{\{{\bm x},{\bm p},m,q\}}\equiv q \,\delta^3({\bm x}-{\bm y})\,.
\ee
are the contributions of a particle occupying the state labeled by $A$ to the local value of the conserved quantities labeled by $a$. 
Note that the repeated capital-letter (hadronic) indices on the right-hand side of Eq.~(\ref{Eq:cons}) imply summation over all hadronic quantum numbers and integration over the phase space variables for the hadrons. The local conservation law also relates the hydrodynamic and hadronic correlators:
\begin{equation}
\label{Eq:constraint}
    H_{a_1\dots a_k} = G_{A_1\dots A_k}  P^{A_1}_{a_1}\dots P^{A_k}_{a_k}\,.
\end{equation}
However, for the freeze-out, we are interested in the inverse problem; namely calculating the hadronic correlators with the knowledge of the hydrodynamic correlators, which is not trivial since Eq.~\eqref{Eq:constraint} is not invertible. This is because the hadronic distribution depends on many variables such as the momenta, charges and masses of hadrons, whereas the hydrodynamic densities only depend on position. Maximum-entropy method allows one to identify the most likely hadronic distribution under the constraints given by Eq.\eqref{Eq:constraint}. 

In general, even in non-interacting hadron gas there is a nonzero contribution of \textit{local} fluctuations to correlation functions. We are, however, interested in ``genuine" correlations which result from nontrivial interactions, critical phenomena and so on. Therefore we subtract the local fluctuations contributions of the ideal hadron gas from the correlation functions. We denote the hadron resonance gas contributions to the correlators with a bar. For example the HRG contribution to the two point function is
\be
\bar{G}_{AB}=\langle\delta f_A\delta f_B\rangle_{\rm connected}^{\text{HRG}}\,.
\ee
We can also calculate the corresponding contribution to the hydrodynamic correlators via the conservation laws given in Eq.\eqref{Eq:constraint}. For example, the HRG contribution to the hydrodynamic two point function is
\be
\label{Eq:BarH}
\bar{H}_{ab}=\bar G_{AB} P^A_{a}P^{B}_b\,.
\ee

Ref.~\cite{\meref}, introduced novel correlation measures, referred to as the \textit{irreducible relative correlators} (IRC), which quantify the non-trivial correlations, irreducible to lower order correlations, relative to a reference distribution, in this case the uncorrelated hadron resonance gas model. The hydrodynamic IRCs are denoted by $\widehat{\Delta} H$ while the irreducible correlators of particle multiplicities are denoted by $\widehat{\Delta} G$. 
In general the hydrodynamic IRCs, $\widehat{\Delta}H$s, are related to the hydrodynamic correlators after the HRG subtraction, i.e.  $H-\bar H$, via some algebraic relations discussed in detail in \cite{\meref} and \cite{\meappref}.
For the IRCs, we use the same compact notation for sub(super)scripts that we introduced for raw correlators, $H$ and $G$. 
Finally, according to the maximum-entropy freeze-out prescription, the IRCs of particle multiplicities are directly determined by the corresponding irreducible hydrodynamic correlators, as given by the relation:
\be\label{Eq:IRCGAsToIRCH}
	\hat{\Delta} G_{A_1A_2\dots A_k}=\hat{\Delta}H_{a_1a_2\dots a_k} P^{a_1}_{A_1} P^{a_2}_{A_2}\dots P^{a_n}_{A_k}\ ,
	\ee	
 where, $P_{A}^{a}$ is given by the expression 
\begin{equation} \label{Eq:P-lowerA-uppera-defn}   
    P^{a}_{A}\equiv (\bar{H}^{-1})^{ab}P^B_b\bar{G}_{BA}\,,  
\end{equation}
Just like before, the repeated small-letter (hydrodynamic) indices on the right-hand side of Eqs.~(\ref{Eq:IRCGAsToIRCH}) and (\ref{Eq:P-lowerA-uppera-defn}) imply both summation over the hydrodynamic variables, energy density ($\epsilon$) and baryon density ($n$), as well as integration over spatial coordinates. 

Integrating $k^{th}$ order hadron gas IRC,  $\widehat{\Delta }G$, over the phase space (within the detector acceptance range) for the protons and normalizing by the mean multiplicity for protons in the same phase space, we get the normalized factorial cumulant for proton multiplicity denoted by $\widehat{\Delta}\omega_{kp}$. These observables are measured experimentally.

The above discussion is general and does not require the system to be in thermal equilibrium. If we further assume thermal equilibrium, all hydrodynamic correlators can be determined from the equation of state. More precisely, the connected hydrodynamic correlators of energy density, denoted by $\e$, and baryon density, denoted by $n$, in equilibrium follow from differentiating the pressure as in
\be
\label{Eq:ExplicitCorrelationsAll}
H_{(j)\e(k-j)n}\equiv 
\langle (\delta\epsilon)^{j}(\delta n)^{k-j} \rangle^{\rm equilibrium}_{\rm connected} &=&
\frac{(-1)^{j}}{V^{k-1}}\frac{\partial^{k}(\beta P)}{\partial\beta^{j}\partial\hat{\mu}^{k-j}}\,.
\ee
Here $V$ is the volume of a hydrodynamic cell, $\beta=1/T$, and $\hat{\mu}\equiv\mu/T$. Notice that we have introduced a new notation for hydrodynamic correlators. In equilibrium, the densities are homogeneous and fluctuations in different hydrodynamic cells are uncorrelated, therefore we drop the spatial dependence in the subscripts to keep our notation simple, and only keep track of how many $\delta \e$s and $\delta n$s there are in the correlator denoted by the integers in parentheses. Due to homogeneity any $k$-point (connected) correlator has an overall factor of $1/V^{k-1}$. 
Since pressure is the generating functional for correlators, we can subtract the contribution of the reference distribution, which we identify with the hadron resonance gas, directly from the pressure, 
\begin{equation}
    P(\mu,T)=\bar P(\mu,T)+\Delta P(\mu,T)\,,
    \label{Eq:Pdecompose}
\end{equation}
where $\bar P(\mu,T)$ denotes the pressure of the hadron resonance gas, $\bar P=P_{\rm HRG}$. As mentioned earlier, our main focus is the deviations from the pressure of the hadron resonance gas which we denote as $\Delta P$. Following the decomposition of the pressure given in Eq.~\eqref{Eq:Pdecompose}, the hydrodynamic correlators $H_{(j)\epsilon(k-j)n}$ can be decomposed into hadron-resonance gas part plus everything else as follows:
\be
\label{Eq:Hdecompose}
H_{(j)\epsilon(k-j)n}=\bar{H}_{(j)\epsilon(k-j)n}+\Delta H_{(j)\epsilon(k-j)n}\,.
\ee
Finally, as mentioned earlier, $\Delta H$s are related to their irreducible counterparts, $\widehat{\Delta} H$s, via algebraic relations given in Refs. \cite{\meref,\meappref}. The HRG model that we employ to evaluate $\bar{P}_{\rm HRG}$ includes all the hadron species from the SMASH hadronic transport framework \cite{SanMartin:2023zhv}. This list includes resonances, with masses up to 3.2 GeV, in order to agree with the lattice QCD EoS.

\subsection{Equation of State near QCD critical point}
\label{Sec:EoS}

Having established the connection between hydrodynamic and hadronic fluctuations, our next task is to calculate the hydrodynamic correlators in the vicinity of the critical point. In general, as discussed in the previous section, the connected hydrodynamic correlators in equilibrium $H$ follow directly from the EoS via Eq.~\eqref{Eq:ExplicitCorrelationsAll}). 
In the vicinity of the critical point, the major contributions to baryon number susceptibility and other higher order correlators will come from the leading singularity of the pressure. In this paper we will focus on the contribution to $\Delta P$ that comes from $P^{\rm singular}$ given by Eq. \eqref{Eq:Psing},
\begin{equation}
   \Delta P= P^{\text{singular}}(\mu,T)=-T^{4}_c\, G_{\rm{Ising}}(r(\mu,T)\, h(\mu,T))
\end{equation}

Furthermore, as elaborated in the Appendix of Ref.~(\cite{Karthein:2025hvl}), the leading contribution to the factorial cumulants of proton multiplicity come from the critical part of the \textit{baryon density correlators}, $\Delta H_{(k)n}$. These are the correlators we focus on in this section. 
It is useful to work with the correlators that are dimensionless and independent of the volume factors. For this reason we define the rescaled correlator, $\Delta H_{kn}$, as 
\begin{equation}
    \Delta H_{kn}\equiv\frac{V^{k-1}}{T_c^3}\Delta H_{(k)n},
\end{equation}
such that
\begin{equation}
    \Delta H_{kn}=\frac{T_c}{T}\frac{\partial^kG_{\rm Ising}}{\partial\hat\mu^k}\,.
\end{equation}
 The derivatives with respect to the QCD variables have to be converted to the derivatives with respect to the Ising variables. For that purpose we use a map similar to the one introduced in Ref.~\cite{Kahangirwe:2024cny}, which is an improved version of the linear map in Eq.\eqref{Eq:map-linear}. This map is linear in $T-T_c$ and $\mu^{2}-\mu^2_c$:
\be
\label{Eq:map-quad}
h(\mu,T)=-\frac{\Delta T^{'}\cos\alpha_1}{ T_c w \sin\alpha_{12}},\quad
r(\mu,T)= \frac{1}{T_c  w\rho}\left[\frac{\Delta T^{'}\cos\alpha_2}{\sin\alpha_{12}}-\frac{\mu^2-\mu^2_c}{2\mu_c \cos\alpha_1}\right]
\ee
where 
\be
\Delta T^{'}=T-T_c+\tan\alpha_1\frac{\mu^2-\mu_c^2}{2\mu_c}
\ee
Of course to linear order in $T-T_c$ and $\mu-\mu_c$, Eq. \eqref{Eq:map-quad} reduces to the linear map given in Eq. \eqref{Eq:map-linear}. The quadratic dependence on $\mu^2$ reproduces the analytic behavior of the equation of state around $\mu=0$ and, therefore, extrapolates better. The inclusion of the curvature also captures the behavior of the crossover/first order curve more accurately. Using Eq.~(\ref{Eq:map-quad}), we write
\be
\label{Eq:Hknhr}
\Delta H_{kn}= T^{k-1}T_c\sum_{i=0}^{k} \frac{\partial^{k} G_{\text{Ising}}}{\partial h^{i}\partial r^{k-i}} h^{i}_{\mu}r^{k-i}_{\mu}+\dots
\ee
where ``$\dots$'' represents terms involving lower order ($<k$)  derivatives of the Ising Gibbs free energy with respect to $h$ and $r$. As in Eq.\eqref{Eq:map-linear}, $h_\mu=(\partial h/\partial \mu)_{T}$ and $r_\mu=(\partial r/\partial \mu)_{T}$ evaluated at $T=T_c$ and $\mu=\mu_c$. Close to the critical point, the term explicitly written on the RHS of Eq.~(\ref{Eq:Hknhr}) dominates over the terms that are not explicitly mentioned. 
Since in the mapping we have chosen in Eq.~(\ref{Eq:map-quad}), the Ising variables are a quadratic polynomial of $\mu$, the omitted terms are non-zero in our calculation. The effects of these terms become significant away from the critical point and therefore we have retained them. In Section.~(\ref{Sec:Contours}), we show the contour plots for the critical contribution to the connected $k$ point baryon density correlators.

Having obtained the baryon density correlators for the family of equations of state parameterized by different choices of the mapping parameters allowed within the Pad\'e uncertainties, we can calculate the factorial cumulants of proton multiplicity, $\widehat{\Delta}\omega_{kp}$ using the maximum entropy prescription summarized in Section.~(\ref{Sec:ME}).

 Before discussing the specific properties of the proton number cumulants, we point out a general and non-trivial consequence of the critical scaling property.
 The Ising equation of state satisfies the Widom scaling law,
\begin{equation}
G_{\rm Ising}( \lambda^{\frac{1}{\beta\delta}}r , \lambda h)=\lambda^{1+\frac{1}{\delta}}G_{\rm Ising}(r, h)\,.
\label{Eq:Widom}
\end{equation}
From Eq.~\eqref{Eq:map-quad} it follows that for fixed $\mu$ and $T$, a rescaling of $r$ and $h$ is equivalent to a rescaling of $\rho$ and $w$:
\begin{equation}
   r(\mu, T; \lambda^{1-\frac{1}{\beta\delta}}\rho; \lambda^{-1} w) =\lambda^{\frac{1}{\beta\delta}}r(\mu,T;\rho,w);\quad
    h(\mu, T; \lambda^{1-\frac{1}{\beta\delta}}\rho; \lambda^{-1} w) =\lambda h(\mu,T;\rho,w)
\end{equation}
The particular rescaling in Eq.~\eqref{Eq:Widom} is equivalent to $w\rightarrow \lambda^{-1} w$ while keeping $\bar{\rho}\equiv \rho w^{1-\frac{1}{\beta\delta}}$, as defined in Eq.~\eqref{Eq:rho-bar},  fixed. As a result the baryon correlators scale as
\begin{eqnarray}
    \Delta H_{kn}(\mu,T;\bar\rho , \lambda^{-1}w) =\lambda^{1+\frac1\delta}\Delta H_{kn}(\mu,T;\bar\rho,w)\,.
\end{eqnarray}
with the remaining parameters, $\{\mu_c,T_c,\alpha_1,\alpha_2\}$, fixed. The factorial cumulants of proton multiplicities inherit the same rescaling property:
\be
\widehat{\Delta}\omega_{kp}(\mu,T; \bar{\rho},\lambda^{-1} w)=\lambda^{1+\frac{1}{\delta}}\widehat{\Delta}\omega_{kp}(\mu,T; \bar{\rho},w)
\label{Eq:wScaling}
\ee
This means that the qualitative features of $\widehat{\Delta}\omega_{kp}$ such as the locations of the peaks and dips depend only on the parameters $\{\mu_c,T_c, \alpha_1, \alpha_2,\bar{\rho}\}$, whereas $w$ only changes the overall magnitude of the signal without changing its shape. 

Now that we have seen that a rescaling by $w$ will only change the overall magnitude of the observables, $\widehat{\Delta}\omega_{kp}$, we shall study how varying the remaining non-universal mapping parameters, $\{\mu_c,T_c, \alpha_1, \alpha_2,\bar{\rho}\}$ varies the shape of $\widehat{\Delta}\omega_{kp}$ along the freeze-out curve and the location of the characteristic signatures of the critical point, such as the peaks and dips of $\widehat{\Delta}\omega_{kp}$ along the freeze-out curve. This is discussed in Section.~(\ref{Sec:Results}).

\subsection{Freeze-out scenarios}
\label{SubSec:freeze-outScenarios}

\begin{figure}
\center
\includegraphics[scale=0.45]{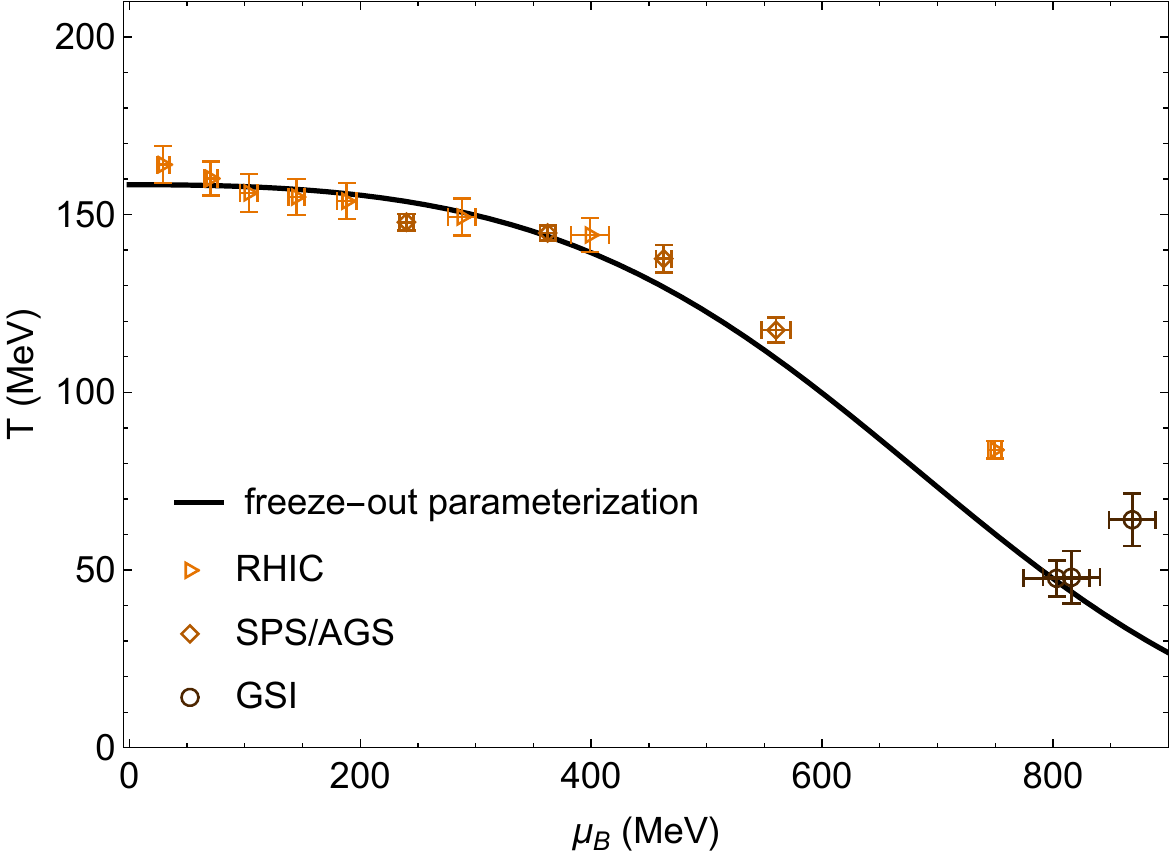}
 \captionof{figure}{A parameterization of the experimental freeze-out curve given in Ref. \cite{Andronic:2017pug} overlaid with different freeze-out data points.}
\label{Fig:fo}
\end{figure}

The final ingredient of our analysis is a parameterization of the chemical freeze-out of hadrons.  The chemical freeze-out points, shown in Fig. \ref{Fig:fo}, are obtained, in a nutshell, by fitting the spectrum of observed hadrons at a given beam energy to thermal yields calculated from the hadron resonance gas. The temperature, chemical potential and volume of the hadron gas are used as fit parameters. As mentioned earlier in Sec \ref{Sec:ME}, this so-called statistical model describes the observed yields of mesons, hadrons even light nuclei, which span seven orders of magnitude, remarkably well \cite{Andronic_2018}. At the same time, its physical interpretation have been questioned for light nuclei \cite{Cai_2019,Cohen:2024wgs}. Nevertheless, at small values of the chemical potential, the freeze-out points agree remarkably well with the chiral crossover curve computed on the lattice as seen in Fig. \ref{Fig:pd}. 
In this paper, we shall use a parameterization of the experimental points, denoted by $T_{F}(\mu)$, introduced in Ref. \cite{Andronic:2017pug} .

As elaborated in Sections.~(\ref{Sec:Pade}), improved Pad\'e estimates place constraints on the location of the critical point, i.e the values of $\mu_c$ and $T_c$, the slope of the first-order line at the critical point given by the tangent of angle $\alpha_1$, and the scale factor $\bar{\rho}\equiv \rho w^{1-\frac{1}{\beta\delta}}$. The set of constraints allow for four topologically different scenarios distinguished by the location of the critical point and the slope of the crossover/first-order curve in relation with the freeze-out curve. In particular, the critical point can be above or below the freeze-out curve. We shall name these cases as ``hot" and ``cool" critical points respectively. Independently, for both of these two cases, the crossover curve can either cross the freeze-out curve or not. We shall name these subcases as ``with crossing'' or ``without crossing'', respectively. Therefore the four distinct scenarios are:

\textbf{Hot critical point without crossing (H)} : This is the most commonly discussed case in literature, where the the freeze-out temperature, $T_{f}(\mu)$ remains below cross-over temperature, $T_{\text{co}}(\mu)$ for all $\mu\leq \mu_c$, on the cross-over side of the phase diagram.

\textbf{Hot critical point with crossing (HX)} : In this scenario, the critical temperature satisfies $T_c>T_f(\mu_c)$ but the freeze-out curve intersects the cross-over line at some $\mu=\mu_{\times}<\mu_c$ . Specifically, $T_{f}(\mu)>T_{\text{co}}(\mu)$, for $0\leq\mu<\mu_{\times}<\mu_c$ and $T_{f}(\mu)<T_{\text{co}}(\mu)$, for $\mu_{\times}<\mu\leq\mu_c$. 

\textbf{Cool critical point with crossing (CX)} : The third scenario which we shall consider has $T_c<T_f(\mu_c)$, but for some $\mu=\mu_{\times}<\mu_c$ the freeze-out curve intersects with the cross-over curve. In this case, $T_{f}(\mu)<T_{\text{co}}(\mu)$, for $0
\leq\mu<\mu_{\times}$ and $T_{f}(\mu)>T_{\text{co}}(\mu)$,for $\mu_{\times} <\mu\leq \mu_c$. 

\textbf{Cool critical point without crossing (C)} : In the last scenario we consider, the freeze-out temperature lies above than the cross-over temperature for all values of $\mu\leq\mu_c$, i.e, $T_{f}(\mu)>T_{\text{co}}(\mu)$ , $\forall\, 0\leq \mu\leq \mu_c$ on the cross-over side.

The notion of the cool critical point, where the critical temperature is less than the freeze-out temperature, might be seen as conceptually problematic. In fact, this line of reasoning was used to rule out any critical point position estimates which fall below freeze-out \cite{Lysenko_2025}. In this paper we will instead explore the implications of considering such a possibility, and demonstrate that it leads to distinct experimentally observable properties of the proton multiplicity cumulants. To justify considering such a scenario, without considering the details of the freeze-out physics, one could say that the freezeout occurs somewhere on the {\em diffuse\/} boundary between the QGP and HRG --- the deconfinement crossover transition region discussed in Sec \ref{Sec:generalities}. The critical point also occurs somewhere on this diffuse boundary. The values of the chiral condensate, the energy density, and the Polyakov loop slightly below and slightly above the crossover line are not dramatically different. Even in the first order region, close to the critical point, the discontinuity in these observables would remain small. Therefore, if the freezeout can occur just below the crossover line or the critical point, it can also occur just above, still within that diffuse deconfinement and chiral transition region. 

Even though the average values of densities, condensates, or the Polyakov loop vary little in the critical region, the susceptibilities, and thus fluctuations, undergo more dramatic changes.
Therefore, these four different scenarios can show qualitatively different characteristic signatures if the crossing happens in the critical region where the leading singular behavior from the Ising like critical point dominates. How the placement of the freeze-out line relative to the crossover line affects the observables is discussed in Section.~(\ref{Sec:Results}). 

\section{Baryon density cumulants}
\label{Sec:Contours}
\begin{figure}
\center
\includegraphics[scale=0.5]{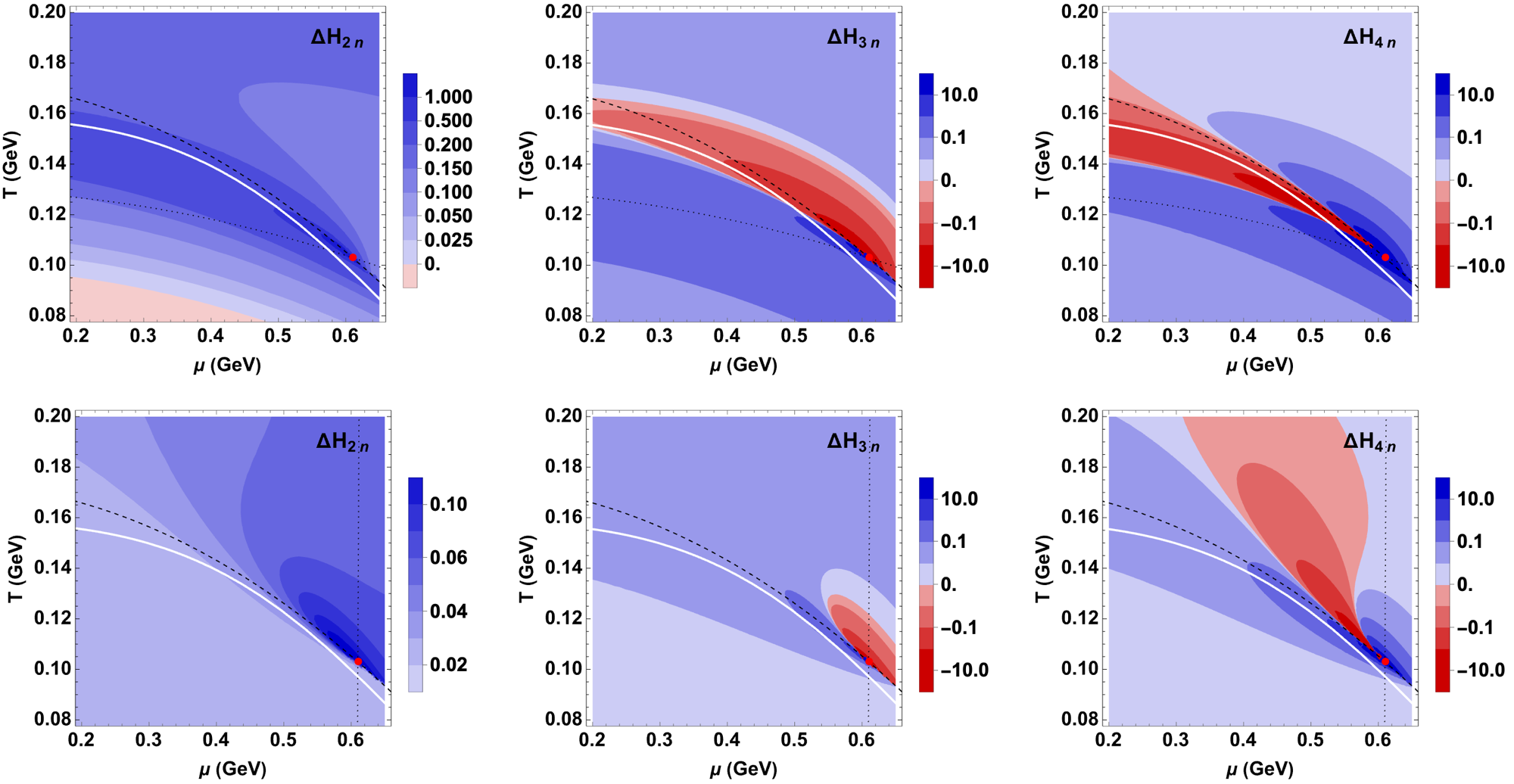}
 \captionof{figure}{An example of scenario H (hot critical point without crossing). $\Delta H_{2n}, \, \Delta H_{3n},\, \Delta H_{4n}$ for
$\mu_c = 610\,\text{MeV}$, $T_c = 103\,\text{MeV}$, $\alpha_1 = 13^{\circ}$, $\bar{\rho}=\bar{\rho}_{\text{max}}$, and $w = 5$. Top row: $\alpha_2=5^{\circ}$, bottom row: $\alpha_2=-89^{\circ}$. The dashed black, the dotted black and the white curves represent the $r$ axis (the cross-over curve), the $h$ axis and the freeze-out curve respectively. 
}
 \label{Fig:HydroHotWithoutCrossing}
 \end{figure}

\begin{figure}
\center
\includegraphics[scale=0.5]{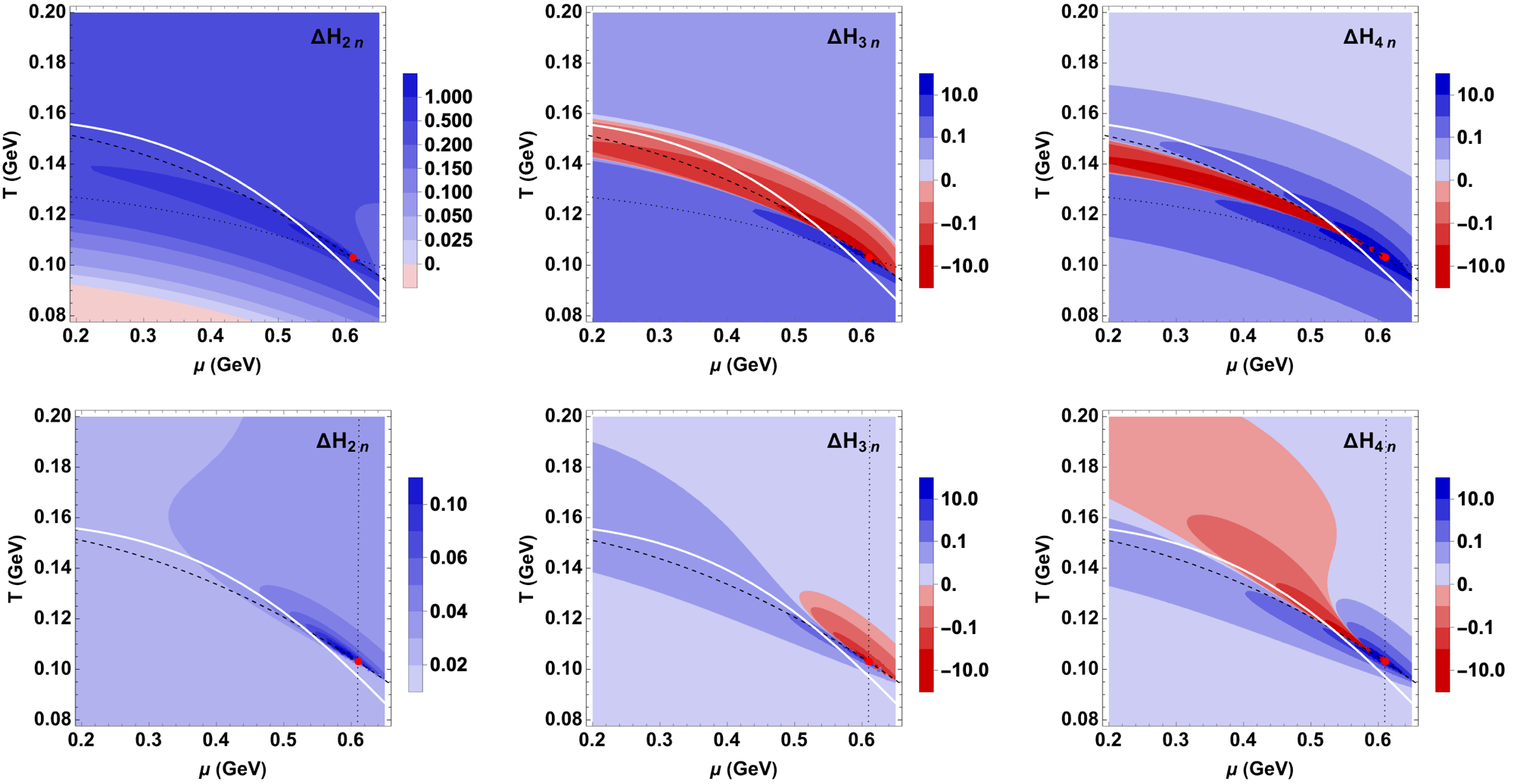}
 \captionof{figure}{An example of scenario HX (hot critical point with crossing). $\Delta H_{2n}, \, \Delta H_{3n},\, \Delta H_{4n}$ for 
$\mu_c = 610\,\text{MeV}$, $T_c = 103\,\text{MeV}$, 
$\alpha_1 = 10^{\circ}$, $\bar{\rho}=\bar{\rho}_{\text{max}}$, and $w = 5$. Top row: $\alpha_2=5^{\circ}$, bottom row: $\alpha_2=-89^{\circ}$. The dashed, dotted and thick lines are the same as in Fig.~(\ref{Fig:HydroHotWithoutCrossing}).  
}
 \label{Fig:HydroHotWithCrossing}
 \end{figure}

\begin{figure}
\center
\includegraphics[scale=0.5]{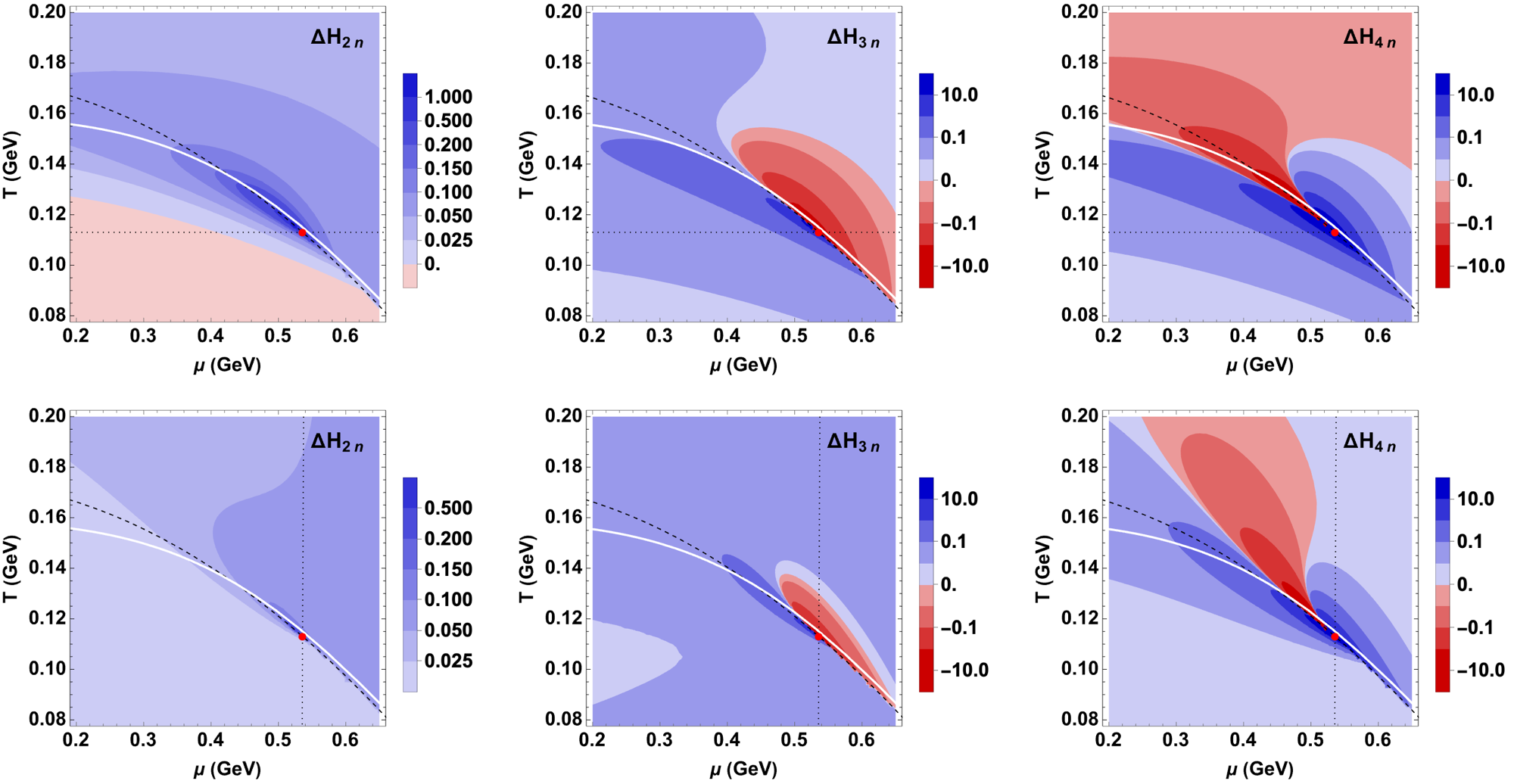}
 \captionof{figure}{An example of scenario CX (cool critical point with crossing). $\Delta H_{2n}, \, \Delta H_{3n},\, \Delta H_{4n}$ for
$\mu_c = 536\,\text{MeV}$, $T_c = 113\,\text{MeV}$, $\alpha_1 = 13^{\circ}$, $\bar{\rho}=\bar{\rho}_{\text{max}}$, and $w = 5$. Top row: $\alpha_2=0^{\circ}$, bottom row: $\alpha_2=-89^{\circ}$. The $r$ and $h$ axes and the freeze-out curve are represented by the same line styles as in Fig.(\ref{Fig:HydroHotWithoutCrossing}). }
 \label{Fig:HydroColdWithCrossing}
 \end{figure}

\begin{figure}
\center
\includegraphics[scale=0.5]{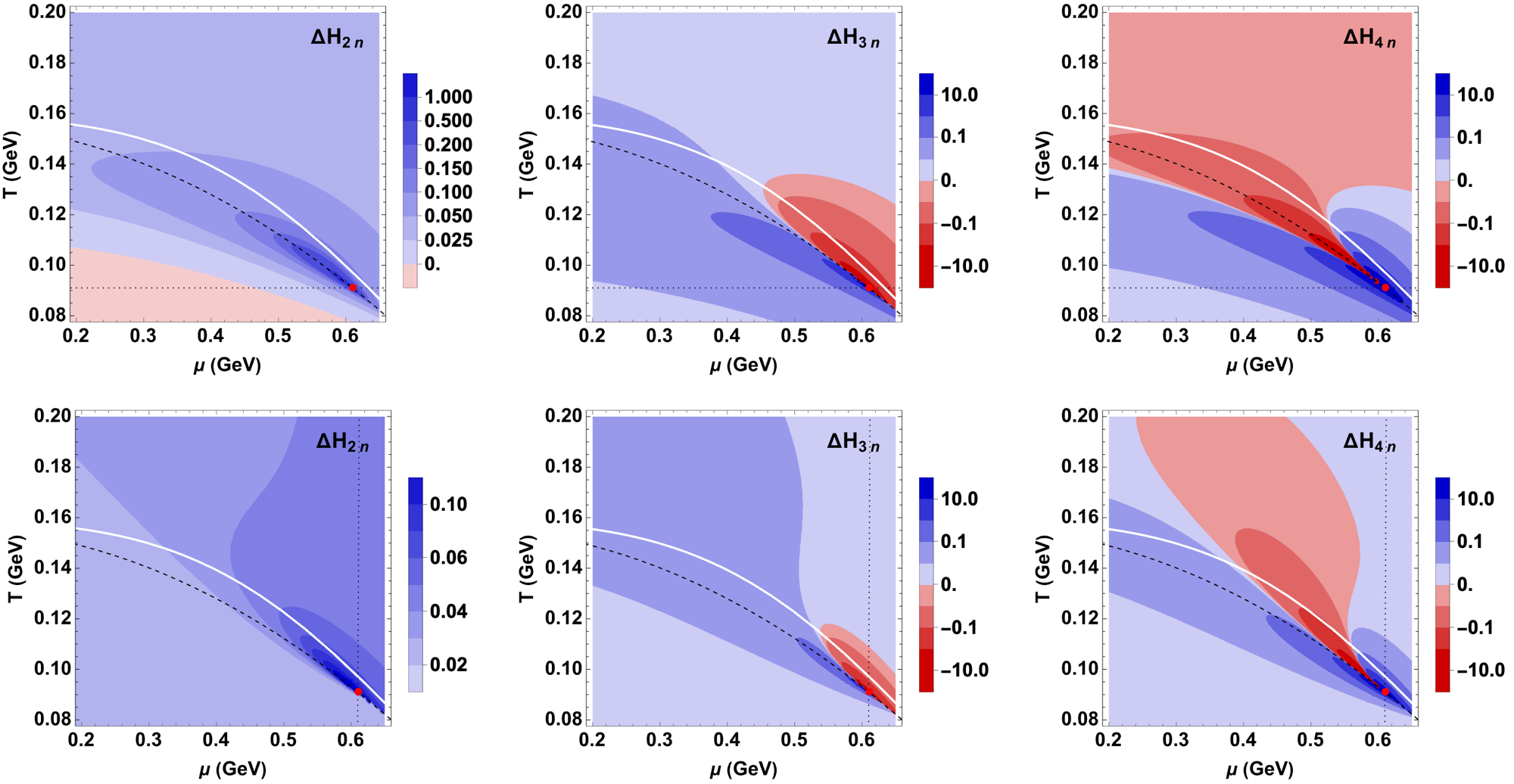}
 \captionof{figure}{An example of scenario C (cool critical point without crossing). $\Delta H_{2n}, \, \Delta H_{3n},\, \Delta H_{4n}$ for
$\mu_c = 610\,\text{MeV}$, $T_c = 91\,\text{MeV}$, $\alpha_1 = 12^{\circ}$, $\bar{\rho}=\bar{\rho}_{\text{max}}$, and $w = 5$. Top row: $\alpha_2=0^{\circ}$, bottom row: $\alpha_2=-89^{\circ}$ . Dashed, dotted and white lines are defined as in Fig.~(\ref{Fig:HydroHotWithoutCrossing}). 
} 
 \label{Fig:HydroColdWithoutCrossing}
 \end{figure}

\begin{figure}
\center
\includegraphics[scale=0.5]{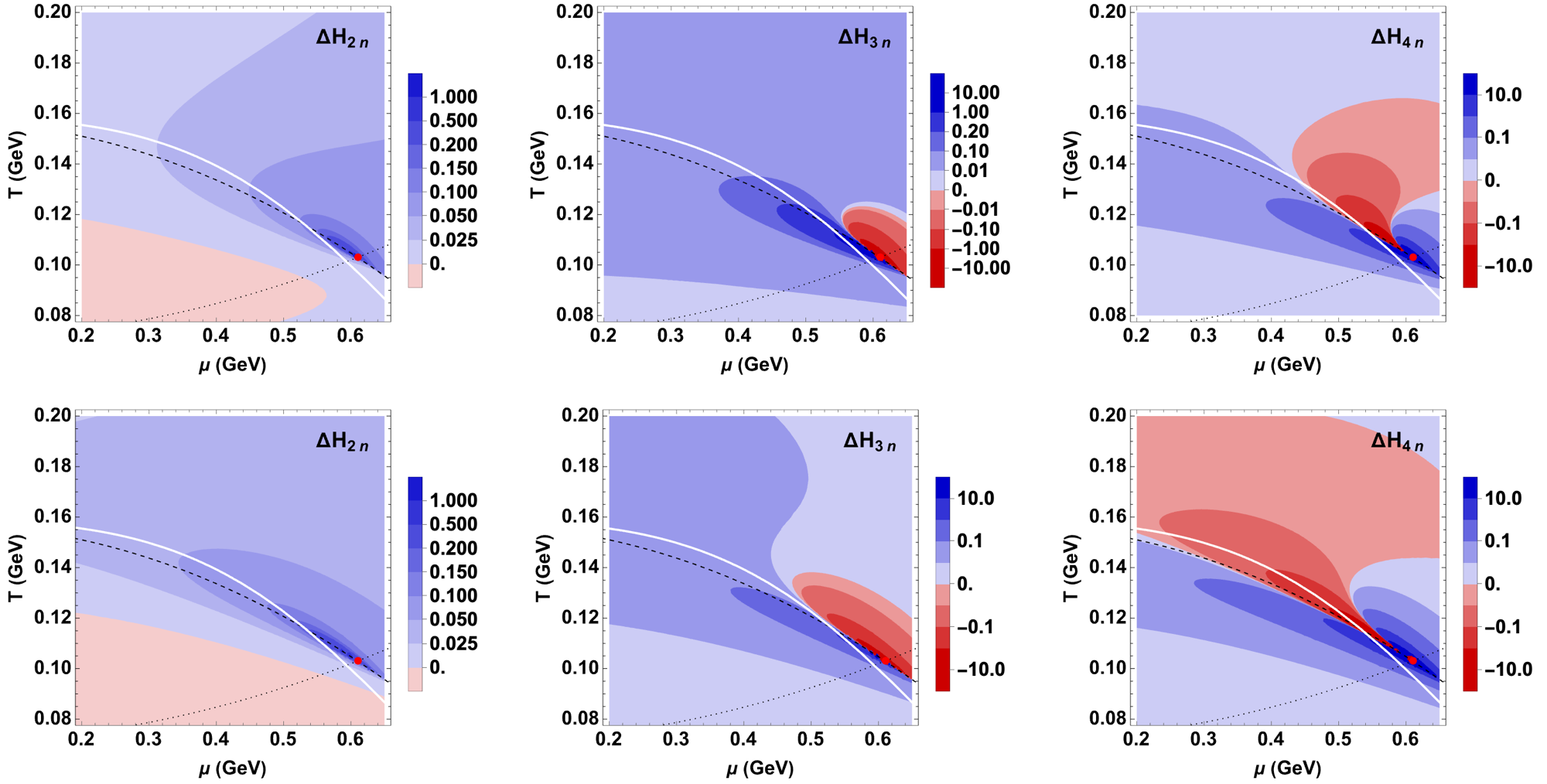}
 \captionof{figure}{An illustration of the effect of varying $\bar\rho$. $\Delta H_{2n}, \, \Delta H_{3n},\, \Delta H_{4n}$ for 
$\mu_c = 610\,\text{MeV}$, $T_c = 103\,\text{MeV}$, $\alpha_1 = 10^{\circ}$, $\alpha_2 = -6^{\circ}$, and $w = 5$.  Top row: $\bar{\rho}=\bar{\rho}_{\text{min}}$,  bottom row: $\bar{\rho}=\bar{\rho}_{\text{max}}$. Dashed, dotted and white lines are defined as in Fig.~(\ref{Fig:HydroHotWithoutCrossing}). Increasing $\bar\rho$ pushes signatures further away from the critical point.}
 \label{Fig:HydroHotWithCrossingRhoDependence}
 \end{figure}

In this Section we  present the contour plots of the critical contribution to cumulants of baryon density for the different choices of the mapping parameters tabulated in Table. \ref{table:parameters-Pade}. These mapping parameters are chosen within the range indicated by the Pad\'e calculations. In Figs.~(\ref{Fig:HydroHotWithoutCrossing},\ref{Fig:HydroHotWithCrossing},\ref{Fig:HydroColdWithCrossing},\ref{Fig:HydroColdWithoutCrossing},\ref{Fig:HydroHotWithCrossingRhoDependence}), we have plotted the contours of $\Delta H_{2n}, \,\Delta H_{3n}$ and $\Delta H_{4n}$ in the left, middle and right panels respectively. Figs.~(\ref{Fig:HydroHotWithoutCrossing},\ref{Fig:HydroHotWithCrossing}) show the contours of $\Delta H_{kn}$ for a hot critical point. The top and bottom panels in these plots correspond to an $\alpha_2$ value of $5^{\circ}$ and $-89^{\circ}$ respectively. Figs.~(\ref{Fig:HydroColdWithCrossing},\ref{Fig:HydroColdWithoutCrossing}) show the contours of $\Delta H_{kn}$ for a cold critical point. The top and bottom panels in these plots correspond to an $\alpha_2$ value of $0^{\circ}$ and $-89^{\circ}$ respectively. The choices of $\alpha_1$ have been made so as to represent a ``hot critical point without crossing" scenario H in Fig.~(\ref{Fig:HydroHotWithoutCrossing}),  a ``hot critical point with crossing" scenario HX in Fig.~(\ref{Fig:HydroHotWithCrossing}),  a ``cool critical point with crossing" scenario CX in Fig.~(\ref{Fig:HydroColdWithCrossing}), and a ``cool critical point without crossing" scenario C in Fig.~(\ref{Fig:HydroColdWithoutCrossing}). In Fig.~(\ref{Fig:HydroHotWithCrossingRhoDependence}) we take one of these choices(``hot critical point with crossing") to demonstrate how the contour plots change as $\bar{\rho}$ increases. A change in the location of the critical point produces a linear displacement in $\mu^2-T$ phase diagram as expected. The value  of angle $\alpha_1$ will play a crucial role in determing the relative placement of the freeze-out curve relative to the cross-over line. In all of these plots, the cross-over ($h=0$) curve, and the freeze-out curve have been shown as black dashed and white solid lines, respectively, allowing the reader to clearly identify which scenario that it corresponds to in the classification discussed in Section.~(\ref{SubSec:freeze-outScenarios}).

Varying $\alpha_1-\alpha_2$ modifies the geometry of the contour plots as well as the labels on the contours, which determine the magnitude of the hydrodynamic correlators, $\Delta H_{kn}$. We summarize some of the important observations here. $\alpha_2=0$ corresponds to a special situation in which the contours of $\Delta H_{kn}$ exhibit certain symmetric properties about the $h=0$ curve (i.e, the cross-over curve) very close to the critical point. In particular, for odd values of $k$, the $\Delta H_{kn}=0$ contour coincides with the $h=0$ near the critical point. This can be seen from the middle plot in the top panels of Fig.~(\ref{Fig:HydroColdWithCrossing},\ref{Fig:HydroColdWithoutCrossing}) that show the contour plots of $\Delta H_{3n}$ for the choice of $\alpha_2=0$ and different choices of the location of the critical points and angle $\alpha_1$. For negative $\alpha_2$, the zero contour of $\Delta H_{3n}$ turns to higher temperatures relative to the cross-over line. This can be seen from the bottom panels of Figs.~(\ref{Fig:HydroColdWithCrossing},\ref{Fig:HydroColdWithoutCrossing}). For positive $\alpha_2$, $\Delta H_{3n}=0$ turns towards lower temperatures relative to the cross-over line. We can see this in the top panels of Fig.~(\ref{Fig:HydroHotWithoutCrossing}) and Fig.~(\ref{Fig:HydroHotWithCrossing}) which correspond to $\alpha_2=5^{\circ}$ for a hot critical point with different values of $\alpha_1$. Increasing $\alpha_1-\alpha_2$, rotates the zero contours of $\Delta H_{4n}$ that pass through the critical point in the clockwise direction. This can be seen in Fig.~(\ref{Fig:HydroHotWithoutCrossing},\ref{Fig:HydroHotWithCrossing},\ref{Fig:HydroColdWithCrossing},\ref{Fig:HydroColdWithoutCrossing}). For $\alpha_2=0$, the contours are symmetric about the cross-over line in the immediate vicinity of the critical point. In addition to modifying the contour geometry, $\alpha_1-\alpha_2$ also has an affect on the overall magnitude of the $\Delta H_{kn}$ at any point on the phase diagram. In each of these plots in Fig.~(\ref{Fig:HydroHotWithoutCrossing},\ref{Fig:HydroHotWithCrossing},\ref{Fig:HydroColdWithCrossing},\ref{Fig:HydroColdWithoutCrossing},\ref{Fig:HydroHotWithCrossingRhoDependence}), $\tan\alpha_2$ is the slope of dotted black curve, which is the $r=0$ curve or in other words, the Ising $h$ axis.

As shown by Eq.~(\ref{Eq:rho-bar-Pade}) and from the plots in Fig.~(\ref{Fig:rhobar-vs-alpha2}), Pad\'e calculations constrains the value of $\bar{\rho}$ for a given choice of $\{\mu_c, T_c, \alpha_1, \alpha_2\}$. Due to the uncertainty in parameter $c_2$, $\bar{\rho}$ is still allowed to vary in a certain band whose minimum and maximum values, denoted by $\bar{\rho}_{\text{min}}$ and $\bar{\rho}_{\text{max}}$ respectively for a given set of $\{\mu_c,  \alpha_1, \alpha_2\}$. Increasing $\bar{\rho}$ stretches out the contours of constant $\Delta H_{kn}$ in both $\Delta \mu^{2}\equiv \mu^2-\mu_c^2$ and $\Delta T=T-T_c$ directions away from the critical point. This can be seen from Fig.~(\ref{Fig:HydroHotWithCrossingRhoDependence}).

\section{Factorial cumulants of proton multiplicity}
\label{Sec:Results}

The critical contributions to correlators involving fluctuations of the baryon density dominate the factorial cumulants of proton multiplicities \cite{Karthein:2025hvl}. Consequently, the qualitative behavior of the factorial cumulants of proton multiplicities along the freeze-out curve can be inferred from the behavior of baryon density correlators.  The behavior of the observables along the freeze-out curve depends sensitively on the curve's orientation and position with respect to the $h$ and $r$ axes of the mapped Ising coordinates on the QCD phase diagram. For each $k^{\text{th}}$ order factorial cumulant, qualitative trends along the freeze-out curve can be anticipated by determining whether the curve traverses regions of positive or negative values of the $k^{\text{th}}$
order connected baryon density correlator as it enters and exits the critical region. This is explained below for the four freeze-out scenarios discussed in Section.~(\ref{SubSec:freeze-outScenarios}) with a representative choice of mapping parameter for each case.

In this Section, we have made four choices for the set of non-universal parameters $\{\mu_c, T_c, \alpha_1\}$ that corresponds to the four different scenarios we discussed in Section.~(\ref{SubSec:freeze-outScenarios}). For each of these choices, we have plotted $\widehat{\Delta}\omega_{kp}\, , \, k=\{2,3,4\}$ for three choices of $\alpha_{2}$ in the range $\{-\pi/2, \alpha_1\}$ and the minimum and maximum values of $\bar{\rho}$ allowed for the set of non-universal parameters according to Fig.~(\ref{Fig:rhobar-vs-alpha2}). One generic conclusion we arrive at is that the location of the critical signatures, such as dips and peaks of $\widehat{\Delta}\omega_{kp}$ is constrained significantly within the uncertainties of $\bar{\rho}$ obtained from the Pad\'e estimates.

\subsection{Scenario H: Hot critical point without crossing}
\begin{figure}
\center
\includegraphics[scale=0.5]{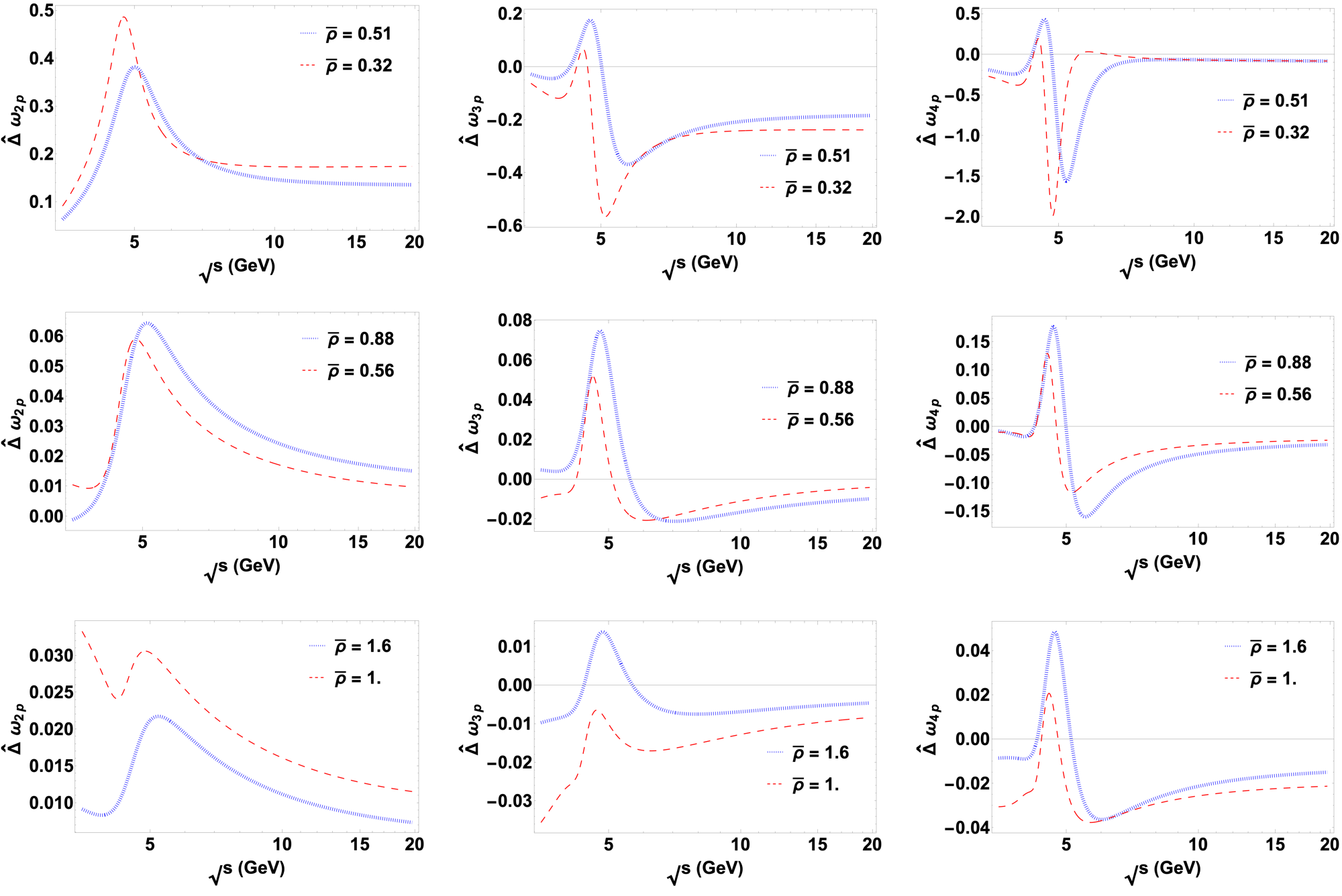}
 \captionof{figure}{Scenario H: $\widehat{\Delta}\omega_{2p}$, $\widehat{\Delta}\omega_{3p}$, and $\widehat{\Delta}\omega_{4p}$ for
$\mu_c = 610\,\text{MeV}$, $T_c = 103\,\text{MeV}$, $\alpha_1 = 13^{\circ}$, and $w = 5$. 
Top row: $\alpha_2 = 5^{\circ}$, middle row: $\alpha_2 = -6^{\circ}$, bottom row: $\alpha_2 = -89^{\circ}$}
 \label{Fig:HadTc103Alpha113}
 \end{figure}

The factorial cumulants of proton multiplicity for a representative choice of $\{\mu_c, T_c, \alpha_1\}$ that corresponds to scenario H is shown in Fig.~(\ref{Fig:HadTc103Alpha113}). We note that $\widehat{\Delta}\omega_{2p}$ has a peak which moves slightly to higher values of $\sqrt{s}$ or lower values of chemical potential with increasing $\bar{\rho}$ and $\alpha_{12}$. We can understand this behavior by looking at the contour plot of $\Delta H_{2n}$ for the same scenario shown in Fig.~(\ref{Fig:HydroHotWithoutCrossing}). At large values of $\alpha_{12}$, a second peak may emerge at a lower collision energy. We can see that this peak appears because of the turning of contours of $\Delta H_{2n}$ as seen in the plot for $\alpha_{2}=-89^{\circ}$ shown in Fig.~(\ref{Fig:HydroHotWithoutCrossing}).

For $\alpha_{2}\leq 0$, $\widehat{\Delta}\omega_{3p}$ exhibits a pronounced peak as the freeze-out curve gets closer to the critical point. For scenario H with $\alpha_2\leq 0^{\circ}$, the freeze-out curve remains entirely below the $\Delta H_{3n}=0$ as shown in Fig.~(\ref{Fig:HydroHotWithoutCrossing}). Consequently, the point on the cross-over side along the freeze-out curve that lies closest to the critical point, where $|\Delta H_{3n}|$ takes it maximum value along the freeze-out curve lies within a region where $\Delta H_{3n}\geq 0$. This ensures that the characteristic critical signature for this  scenario is a peak. The shallow dip in this case is due to the subtraction of the term proportional to $\Delta H_{2n}$ in the definition of $\widehat{\Delta}H_{3n}$\cite{Pradeep:2022eil,Karthein:2025hvl}. If $\alpha_{2}>0$, the freeze-out curve can pass through a sector where $\Delta H_{3n}<0$, before reaching the point of closest approach to the critical point. In this case the peak could be preceded by a  dip at lower values of freeze-out chemical potential, or equivalently higher values of collision energy. This is also seen in Fig.~(\ref{Fig:HadTc103Alpha113}) where the peak grows with increasing $\alpha_{12}$ whereas the dip becomes more pronounced for smaller values of $\alpha_{12}$. 

Increasing $\bar{\rho}$ displaces the peak and dip to higher values of collision energy by a small amount. This can be understood from the characteristic stretching of the contours of $\Delta H_{kn}$, shown in Fig.~(\ref{Fig:HydroHotWithCrossingRhoDependence}) along $\Delta \mu^2$ and $\Delta T$ directions as the $\bar{\rho}$ is increased from $\bar{\rho}_{\text{min}}$ to $\bar{\rho}_{\text{max}}$. This qualitative behavior with $\bar{\rho}$ remains the same across all the scenarios, as can be seen from the plots in this Section.

$\widehat{\Delta}\omega_{4p}$ for this scenario exhibits a dip followed by a peak as the collision energy is lowered. The relative prominence of the peak increases with $\alpha_{12}$ as reflected in the growing peak to dip ratio. As in the case of $\widehat{\Delta} \omega_{3p}$, increasing $\bar{\rho}$ displaces the peak and dip to higher values of collision energy by a small amount.

\subsection{Scenario HX: Hot critical point with crossing}
 \begin{figure}
\center
\includegraphics[scale=0.5]{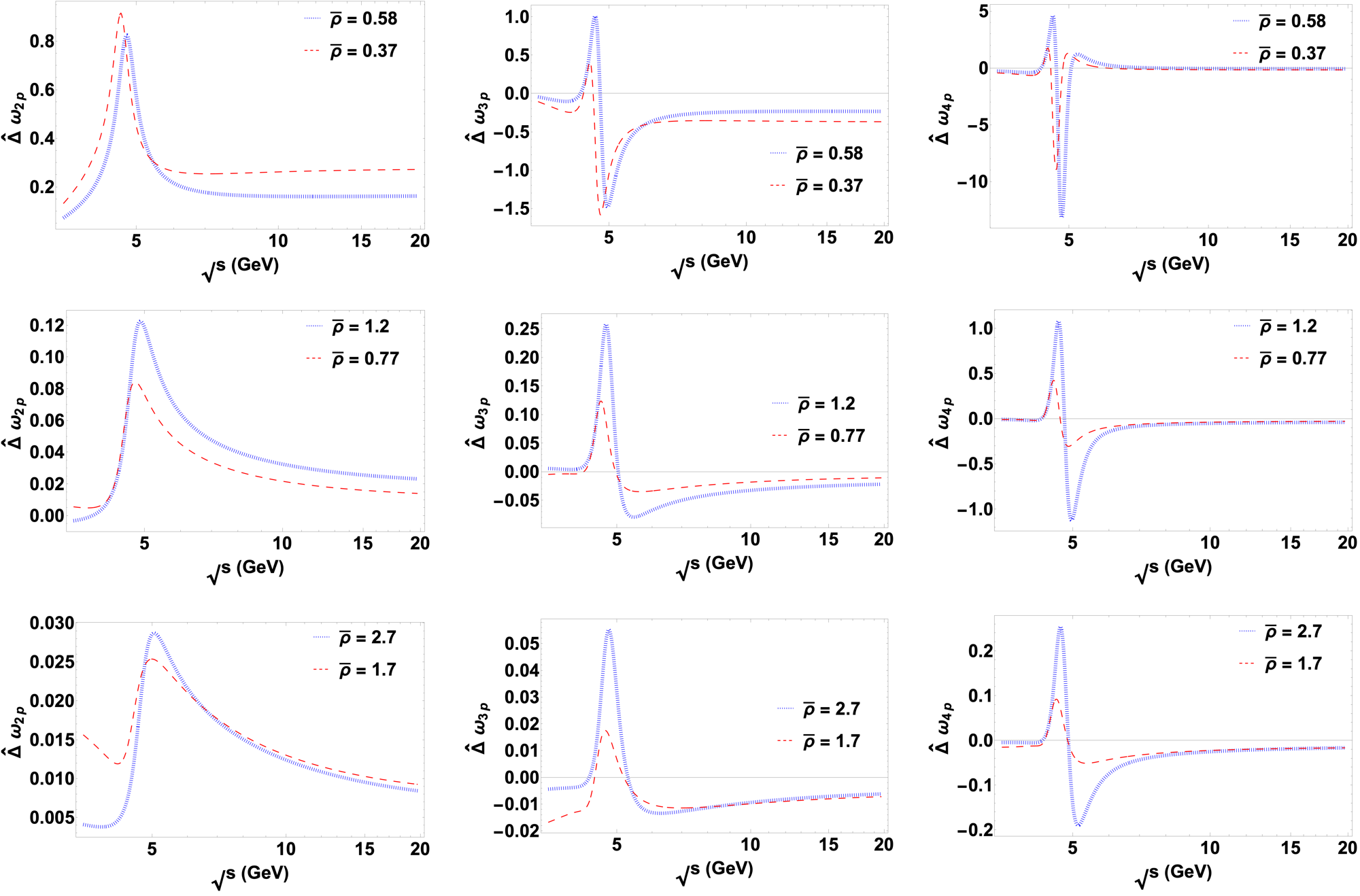}
 \captionof{figure}{Scenario HX: $\widehat{\Delta}\omega_{2p}$, $\widehat{\Delta}\omega_{3p}$, and $\widehat{\Delta}\omega_{4p}$ for
$\mu_c = 610\,\text{MeV}$, $T_c = 103\,\text{MeV}$, $\alpha_1 = 10^{\circ}$, and $w = 5$. 
Top row: $\alpha_2 = 5^{\circ}$, middle row: $\alpha_2 = -6^{\circ}$, bottom row: $\alpha_2 = -89^{\circ}$.}
 \label{Fig:HadTc103Alpha10}
 \end{figure}
The factorial cumulants of proton multiplicity for a representative choice of $\{\mu_c, T_c, \alpha_1\}$ that corresponds to scenario HX is shown in Fig.~(\ref{Fig:HadTc103Alpha10}). The important critical signatures in $\widehat{\Delta}\omega_{kp}$ are similar to scenario H. However, the crossing of the freeze-out curve with the cross-over curve in the vicinity of the critical point, leads to a more sudden change in the behavior of the observables in this scenario. As in all of the scenarios, $\widehat{\Delta}\omega_{2p}$ has a peak which moves slightly to higher values of $\sqrt{s}$ or equivalently, lower values of freeze-out chemical potential with increasing $\bar{\rho}$ and $\alpha_{12}$. 

For lower values of $\alpha_{12}$, the freeze-out curve crosses the $\Delta H_{3n}=0$ contour, as can be noted from Fig.~(\ref{Fig:HydroHotWithCrossing}). This leads to a characteristic dip in $\widehat{\Delta}\omega_{3p}$, marking the critical signature for this scenario. The dip is followed by a peak as the freeze-out curve gets closer to the critical point, i.e as the collision energy is lowered. The peak is more universal property in this scenario as it appears regardless of whether $\Delta H_{3n}=0$ intersects with the freeze-out curve or not. The ratio of the magnitude of peak to dip increases as $\alpha_{12}$ is increased. 

$\widehat{\Delta}\omega_{4p}$ for this scenario exhibits a dip followed by a peak as the collision energy is lowered. The ratio of the magnitude of the peak to dip increases as $\alpha_{12}$ is increased.
For lower values of $\alpha_{12}$, one can see two peaks, one preceding and another one following the dip as the collision energy is varied.

\subsection{Scenario C: Cool critical point without crossing}
 \begin{figure}
\center
\includegraphics[scale=0.5]{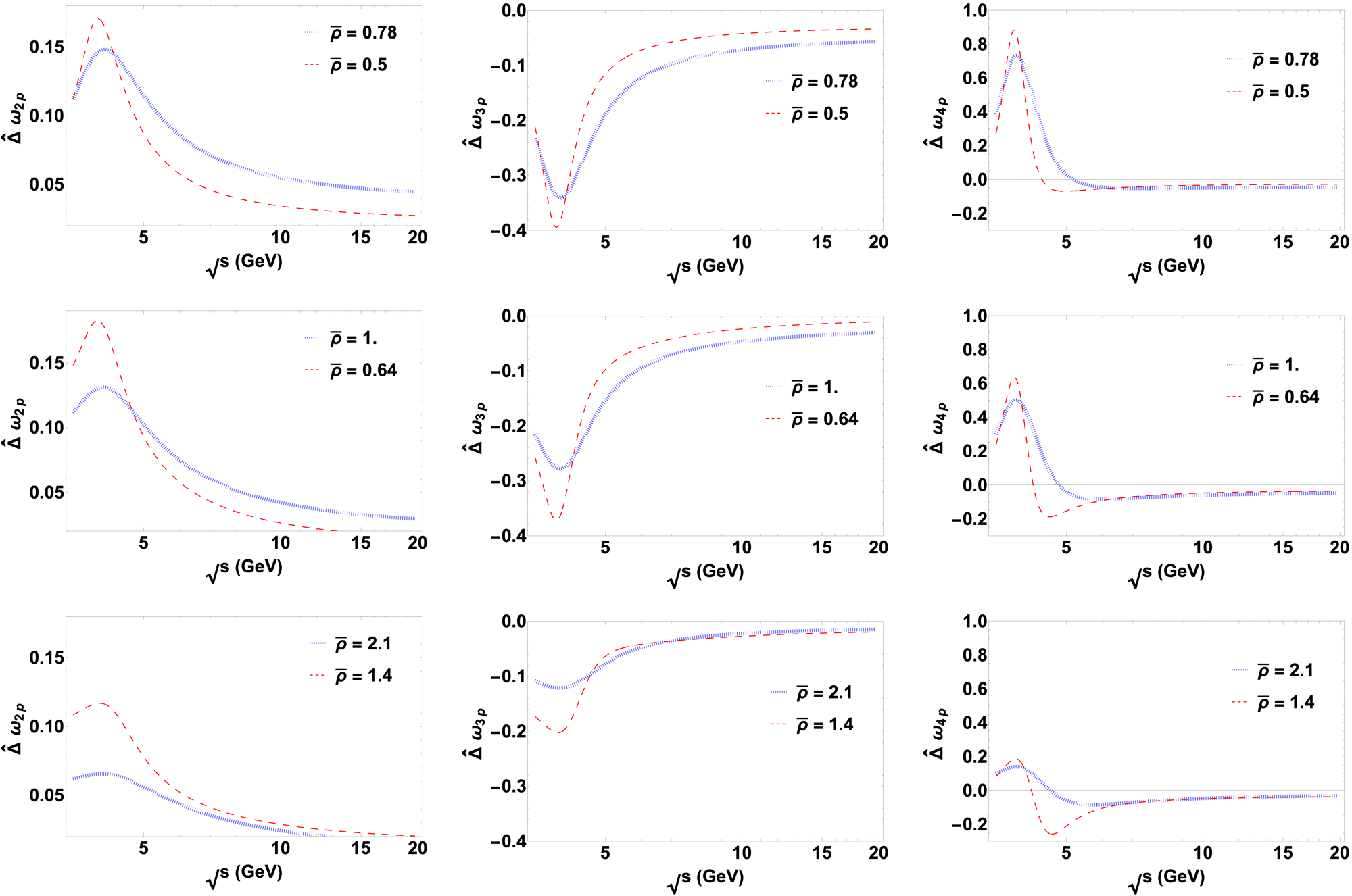}
 \captionof{figure}{Scenario C: $\widehat{\Delta}\omega_{2p}$, $\widehat{\Delta}\omega_{3p}$, and $\widehat{\Delta}\omega_{4p}$ 
for
$\mu_c = 610\,\text{MeV}$, $T_c = 91\,\text{MeV}$, $\alpha_1 = 12^{\circ}$, and $w = 5$. 
Top row: $\alpha_2 = 0^{\circ}$, middle row: $\alpha_2 = -6^{\circ}$, 
bottom row: $\alpha_2 = -89^{\circ}$.}
 \label{Fig:HadTc91Alpha12}
 \end{figure}
 The factorial cumulants of proton multiplicity for a representative choice of $\{\mu_c, T_c, \alpha_1\}$ that corresponds to scenario C is shown in Fig.~(\ref{Fig:HadTc91Alpha12}). We note that $\widehat{\Delta}\omega_{2p}$ has a peak which moves slightly to higher values of $\sqrt{s}$ or lower values of chemical potential with increasing $\bar{\rho}$ and $\alpha_{12}$. 

Investigating the middle panels of Fig.~(\ref{Fig:HydroColdWithoutCrossing}), we can observe that the point of closest approach to the critical point along the freeze-out curve lies in the regime where $\Delta H_{3n}<0$. This leads to a characteristic dip in $\widehat{\Delta}\omega_{3p}$ as the collision energy is lowered. 
The magnitude of the dip falls with increasing $\alpha_{12}$. 

$\widehat{\Delta}\omega_{4p}$ for scenario C exhibits a dip followed by a peak as the collision energy is lowered. The magnitude of the peak becomes more pronounced with increasing $\alpha_{12}$ as reflected in the drop of the ratio of the magnitude of the peak to dip with $\alpha_{12}$. 

\subsection{Scenario CX: Cool critical point with crossing}
  \begin{figure}
\center
\includegraphics[scale=0.5]{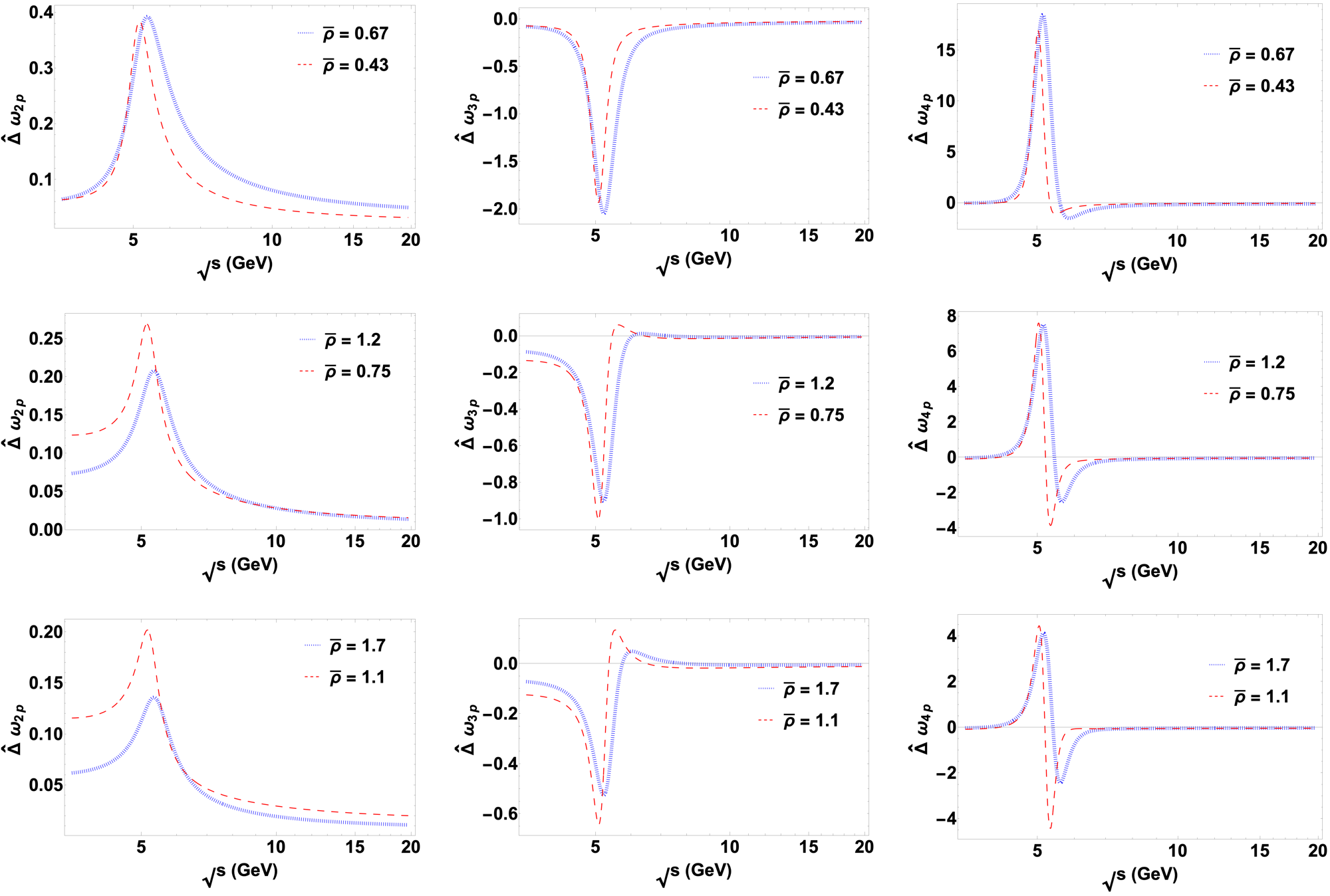}
 \captionof{figure}{Scenario CX: $\widehat{\Delta}\omega_{2p}$, $\widehat{\Delta}\omega_{3p}$, and $\widehat{\Delta}\omega_{4p}$ for
$\mu_c = 536\,\text{MeV}$, $T_c = 113\,\text{MeV}$, $\alpha_1 = 13^{\circ}$, and $w = 5$. 
Top row: $\alpha_2 = 0^{\circ}$, middle row: $\alpha_2 = -20^{\circ}$, bottom row: $\alpha_2 = -89^{\circ}$.}
 \label{Fig:HadTc113Alpha13}
 \end{figure}
The factorial cumulants of proton multiplicity for a representative choice of $\{\mu_c, T_c, \alpha_1\}$ that corresponds to  scenario CX is shown in Fig.~(\ref{Fig:HadTc113Alpha13}). We note that $\widehat{\Delta}\omega_{2p}$ has a peak which moves slightly to higher values of $\sqrt{s}$ or lower values of chemical potential with increasing $\bar{\rho}$ and $\alpha_{12}$. 

For all values of $\alpha_{12}$ and the corresponding range of $\bar{\rho}$ allowed by the uncertainties of the Pad\'e estimation, $\widehat{\Delta}\omega_{3p}$ exhibits a dip as the freeze-out curve gets closer to the critical point. The magnitude of the dip falls with $\alpha_{12}$.  For large $\alpha_{12}$, where the crossing of the freeze-out curve with the $\Delta H_{3n}=0$ contour happens to close to the critical point, a peak occurs at a higher collision energy relative to the position of the dip.  This can be understood from how $\Delta H_{3n}$ behaves along the freeze-out curve as seen from Fig.~(\ref{Fig:HydroColdWithCrossing}).

$\widehat{\Delta}\omega_{4p}$ for scenario CX exhibits a dip followed by a peak as the collision energy is lowered. The ratio of the magnitude of the peak to dip decreases as $\alpha_{12}$ is increased. Increasing $\bar{\rho}$ displaces the peak and dip to higher values of collision energy by a small amount.

 \section{Summary and Discussion}

In this work, we presented a connection between the analytical structure of the QCD equation of state near a conjectured critical point and cumulants of proton multiplicities, the key experimental observables in the search for the QCD critical point.  We utilized the conformal Pad\'e approach to Lee-Yang singularities to constrain the values of the non-universal (QCD specific) parameters such as the location of the  critical point, the slope of the first-order line, and the special combination of the scale parameters, $\bar{\rho}\equiv \rho w^{1-1/\beta\delta}$. We used the Taylor coefficients of pressure obtained by the HotQCD collaboration as our input for the Pad\'e estimation. Within the statistical uncertainties, we identified a region of parameters that shape the singular contribution to the equation of state from which we computed the critical contribution to the connected baryon density correlators. 
In order to demonstrate the features of these correlators, we chose a few exemplary values for the non-universal mapping parameters that are allowed within the uncertainties of Pad\'e resummation to fix the equation of state. The results are shown in Section.~(\ref{Sec:Contours}). Then, by using the maximum entropy freeze-out framework we converted these baryon number cumulants into the factorial cumulants of proton number multiplicities which are presented in Section \ref{Sec:Results}. 
Remarkably, we demonstrated that up to the orientation of the Ising $h$ axis on the QCD phase diagram, determined by the angle $\alpha_2$, the properties of the Lee-Yang singularities fix the shape of the factorial cumulants of proton multiplicities along the freeze-out curve in the critical regime expected if fluctuations reach equilibrium at freezeout. 

Furthermore, by considering the relative position of the critical point and the slope of the crossover/first-order curve with respect to the freeze-out curve determined by the experiments, we identified four topologically different scenarios. We named these scenarios as ``Hot critical point without crossing (H)", ``Hot critical point with crossing (HX)", ``Cool critical point without crossing (C)" and ``Cool critical point with crossing (CX)". Crucially, each scenario leads to experimentally identifiable features in the proton multiplicity cumulants. For example, the non-Gaussian factorial cumulants show more dramatic changes as a function of collision energy when there is ``crossing", an intersection between the freeze-out curve and the cross-over curves. Also, the third cumulant is very sensitive to whether the critical point is hot or cool. For a hot critical point, a peak in $\widehat{\Delta}\omega_{3p}(\mu)$ is the more dominant feature, with the ratio of the magnitude of the peak to dip increasing with $\alpha_{12}\equiv \alpha_1-\alpha_2$. In contrast, for a cool critical point, a dip is the more pronounced feature with a possible appearance of a small peak at higher collision energies for large $\alpha_{12}$ when the freeze-out curve intersects the cross-over curve in the critical region.

The set of exemplary parameters has been chosen to cover a wide range of possibilities. In particular, we chose the largest and the lowest values of $\bar{\rho}$ within $1\, \sigma$ variation from the mean values, and range of $\alpha_2$s between $-90^{\circ}$ to $5^{\circ}$. In addition to the $\alpha_2$ dependence discussed above, investigating the resulting proton multiplicity cumulants, we also conclude that the location of the characteristic signatures such as the dips and peaks in $\widehat{\Delta}\omega_{kp}$ could be significantly constrained using these bounds for $\bar{\rho}$ obtained from the Lee-Yang singularity analysis.  

 It is important to emphasize that the Pad\'e resummations, rely on the extrapolations of a few lowest order Taylor coefficients of the EoS at $\mu=0$ to non-zero and complex values of chemical potential. As such these estimates are subject to uncertainties from the truncation effects, unavailability of the higher order Taylor coefficients as well as pf the lattice data at temperatures lower than $ \sim 130 \, \text{MeV}$. This limitations propagate into our determination of the non-universal parameters as well the predictions for the location of the critical signatures in the observables as a function of collision energy.
 Despite these challenges, we can say that within the uncertainties of the estimates for $\bar{\rho}, \mu_c, T_c$ and $\alpha_1$ obtained from this analysis, the locations of the peak in $\widehat{\Delta}{\omega}_{3p}$ and the dip in $\widehat{\Delta}{\omega}_{4p}$ are substantially displaced from the typical collision energies where the experiment show some indications of non-monotonic features. This discrepancy suggests that the critical point realized in nature is at lower value of chemical potential than predicted by various extrapolations/ calculations or that the non-equilibrium effects play an important role in shifting the location of the critical signatures or both. A more sophisticated Bayesian analysis of the BES data aimed at constraining the likelihood distribution over the non-universal mapping parameters of the equation of state is underway and will be published elsewhere. The present work provides a valuable foundation for this effort, helping to define informed priors in the EoS parameter space and guiding the selection of regions where the likelihood is expected to be maximized.

 In this work we focused only on the singular critical contribution to factorial cumulants assuming local thermodynamic equilibrium. This means that to connect our findings to experimental data in a quantitative fashion, we need to also consider noncritical contributions as well as non-equilibrium effects. We leave this analysis to future work. However, we still were able to identify notably robust features present for all critical point scenarios such as peaks in $\hat\Delta\omega_2$ and $\hat\Delta\omega_4$. 

Finally, we stress that, even though we used a Pad\'e estimation to constrain the QCD EoS, the novel features we identified in this paper do not rely on Pad\'e. Namely, the four topologically different scenarios
and their experimentally distinguishable signatures, as well as the relations between the relations between the Lee-Yang trajectory and the proton multiplicity cumulants in general hold within the assumptions of equilibrium and dominance of the singular part of the EoS.
As a result, it is possible to use other methods, such as functional QCD, or other types of resummations of the equation of state to constrain these parameters. The same conclusions for the proton multiplicity cumulants that we reached here will still hold. 
Alternatively it is also possible to constrain the QCD EoS and the Lee-Yang trajectory by using the experimental data, eventually. Of course, this requires a more quantitative analysis both from the experimental side, such as a more comprehensive understanding of the background, as well as the theory side, such as non-equilibrium dynamics. 

\begin{acknowledgments}
The work of GB is supported by the National Science Foundation CAREER Award PHY-2143149. The work of MP is supported via Ramanujan Fellowship, Project File Number RJF/2025/000614. MP was supported by the DOE grant, DE-FG02-93ER40762 in the initial stages of this project. The work of MS is supported by the U.S. Department of Energy, Office of Science, Office of Nuclear Physics Grant No. DEFG0201ER41195.
\end{acknowledgments}

\appendix

\section{Chiral crossover vs first-order transition}
\label{Sec:appendix-GL}

 We consider the mean-field, Ginzburg-Landau potential
\begin{equation}\label{Eq:Vphi6}
    V(\phi)=-m_q\phi+\frac12a\phi^2+\frac14b\phi^4+\frac16\phi^6\,.
\end{equation}
Here the coefficients $a$ and $b$ are functions of $T$ and $\mu$. In the chiral limit, the tricritical point is located at $a=b=\phi=0$. The second-order and first-order transitions in the chiral limit are given, respectively, as
\begin{equation}
a=\begin{dcases}
    0 \quad {\rm for}\,\, b>0\quad (2^{nd}\,{\rm order})
    \\ 
    \frac3{16}b^2 \quad {\rm for}\,\, b<0\quad (1^{st}\,{\rm order})
\end{dcases}
\end{equation}
They meet at the tricritical point with a continuous slope $da/db$ (zero in this case) but with a discontinuity in the curvature $d^2a/db^2$. 

Away from the chiral limit, for nonzero but small $m_q$, the critical point is located at
\begin{equation}
    \phi=\phi_c\equiv \left(\frac{3m_q}{8}\right)^{1/5},\quad a=a_c\equiv 5\phi_c^4,\quad b=b_c=-\frac{10}{3}\phi_c^2
    \label{Eq:GL-chiral}
\end{equation}
We can parameterize the crossover (defined as the maximum of the chiral susceptibility, $\del \phi/\del m_q$) and the first-order transitions via a curve in the $a$ vs $b$ plane. Similarly to the chiral limit, for $ b>b_c$ we have a crossover and $b<b_c$, a first-order transition. There are two regimes for nonzero quark mass depending on how we approach the critical point. In the immediate vicinity of the critical point where for $|\Delta b|\ll \phi_c^2$ 
where \begin{equation}
    \Delta b\equiv b-b_c
\end{equation}
the domain of universality is given by that of the critical point and the crossover and first-order curves have the following expansions
\begin{equation}
   |\Delta b|\ll\phi_c^2:\quad a-a_c=
    \begin{dcases}
    -\Delta b \,\phi_c^2\left[1-\dfrac{1}{5}\dfrac{\Delta b}{\phi_c^2}+{\cal O}\left(\dfrac{ \Delta b^2}{\phi_c ^4}  \right) \right] & {\rm for}\,\,  \Delta b>0 \quad ({\rm crossover});
    \\
    -\Delta b \,\phi_c^2\left[1-\dfrac{21}{100}\dfrac{\Delta b}{\phi_c^2}+{\cal O}\left(\dfrac{\Delta b^2}{\phi_c ^4}  \right) \right] & {\rm for}\,\,  \Delta b<0\quad (1^{st}\,{\rm order})\,.
    \end{dcases}
    \label{Eq:GL-inside}
\end{equation}
The transition slope $da/db$ is nonzero now, but is still continuous.
The transition line curvature $d^2a/db^2$ is also still discontinuous. 
Outside of this region, when $ |\Delta b|\gg \phi_c^2$ the effects of explicit chiral symmetry breaking and the universality domain is that of the tricritical point.  The expansion  for the crossover and first-order curves in this domain are
\begin{equation}
|\Delta b|\gg\phi_c^2:\quad
     a-a_c=
    \begin{dcases}
    \Delta b^2 \left(\frac{\phi_c^2}{\Delta b}\right)^{5/3} \left[12^{1/3}-5\left(\frac{\phi_c^2}{\Delta b}\right)^{1/3}+{\cal O}\left( \frac{\phi_c ^2}{\Delta b}  \right) \right]  &{\rm for}\,\,  \Delta b>0\quad ({\rm crossover})
    \\ 
    \Delta b^2 \left[\dfrac{3}{16}-\dfrac{5}{4}\dfrac{\phi_c^2}{\Delta b}+{\cal O}\left( \frac{\phi_c ^4}{\Delta b^2}   \right) \right]  &{\rm for}\,\,  \Delta b<0\quad (1^{st}\,{\rm order})\,.
    \end{dcases}
     \label{Eq:GL-outside}
\end{equation}
In the chiral limit, $\phi_c\to0$, the regime of applicability of Eq.~\eqref{Eq:GL-inside} shrinks to a point and the expansion given in Eq.~\eqref{Eq:GL-outside} reduces to Eq.~\eqref{Eq:GL-chiral}.

\begin{figure}[H]
\center
\includegraphics[scale=0.5
]{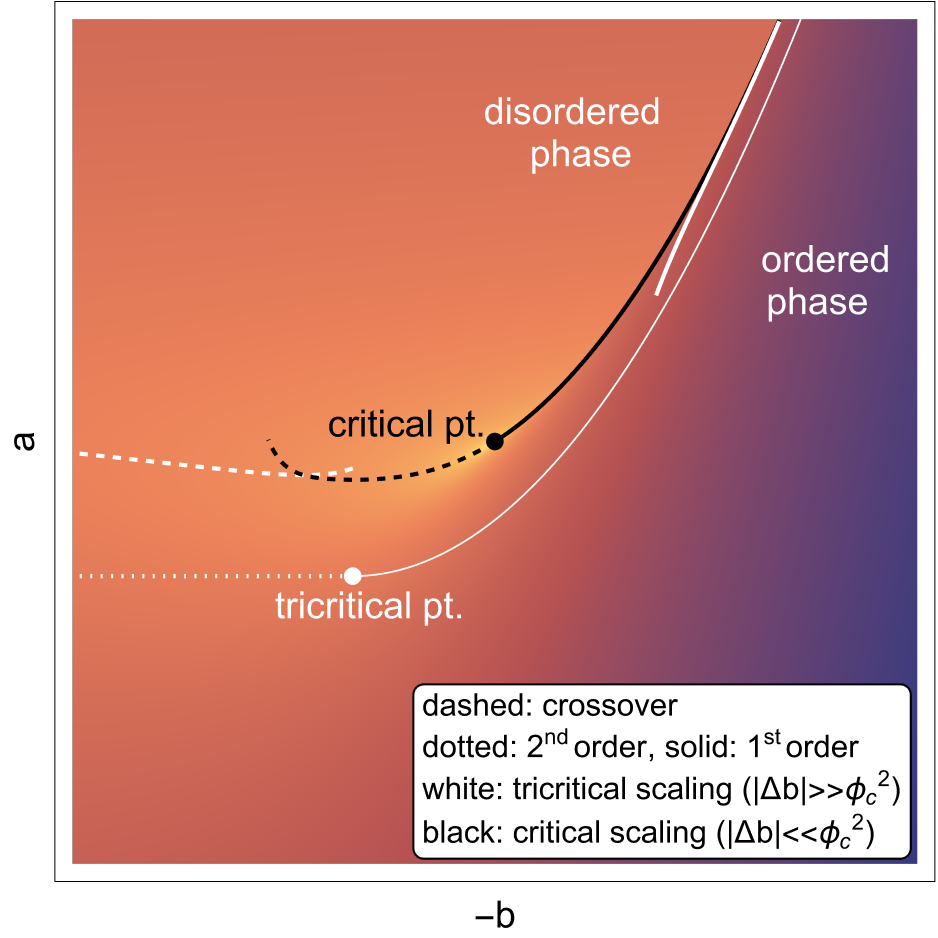}
\captionof{figure}{The phase diagram of the model, given in Eq.\eqref{Eq:Vphi6}, describing transitions in QCD near its tricritical point. The transition between the ordered (chiral symmetry broken) and disordered (chiral symmetry restored) phase in the chiral limit   ($m_q=0$) are shown by white lines meeting at the tricritical point: the solid line denotes first-order transition, the dotted line -- second-order. For $m_q\neq0$ two approximations for the transition line are shown, each obtained by a truncated expansion valid outside of the domain of the universality of the critical point (white) and inside (black). The breakdown of the truncated expansion indicates the range of the validity of the given expansion. The lines on this diagram map onto the QCD phase diagram by a transformation $(a,-b)\to(T,\mu)$ which is approximately linear (with positive Jacobian) near the (tri)critical point. The figure illustrates the abrupt change of the curvature of the transition lines: the curvature increases in the direction of the disordered phase from the second-order/crossover side to the first-order side of the transition.
}
\label{Fig:GL-pd}
\end{figure}

\nocite{*}

\bibliography{apssamp}

\end{document}